\documentclass{aa}  
\usepackage[flushleft]{threeparttable}
\usepackage{multirow} 
\usepackage{natbib,twoopt}
\usepackage{listings}
\usepackage{caption}
\usepackage{appendix}
\usepackage{blindtext} 
\usepackage[breaklinks=true]{hyperref} 
\bibpunct{(}{)}{;}{a}{}{,} 
\makeatletter
\newcommandtwoopt{\citeads}[3][][]{\href{http://adsabs.harvard.edu/abs/#3}%
{\def\hyper@linkstart##1##2{}%
\let\hyper@linkend\@empty\citealp[#1][#2]{#3}}}
\newcommandtwoopt{\citepads}[3][][]{\href{http://adsabs.harvard.edu/abs/#3}%
{\def\hyper@linkstart##1##2{}%
\let\hyper@linkend\@empty\citep[#1][#2]{#3}}}
\newcommandtwoopt{\citetads}[3][][]{\href{http://adsabs.harvard.edu/abs/#3}%
{\def\hyper@linkstart##1##2{}%
\let\hyper@linkend\@empty\citet[#1][#2]{#3}}}
\newcommandtwoopt{\citeyearads}[3][][]%
{\href{http://adsabs.harvard.edu/abs/#3}
{\def\hyper@linkstart##1##2{}%
\let\hyper@linkend\@empty\citeyear[#1][#2]{#3}}}
\makeatother
\usepackage{graphicx}
\usepackage{txfonts}
\begin{document} 

   \title{Revealing the dust grain size in the inner envelope of the Class I protostar Per-emb-50 \thanks{Based on observations carried out under project number S16AT with the IRAM NOEMA Interferometer. IRAM is supported by INSU/CNRS (France), MPG (Germany) and IGN (Spain).}}


   \author{C. Agurto-Gangas
          \inst{1} \and      
          J.E. Pineda\inst{1}\and
          L. Sz\H{u}cs\inst{1}\and
          L. Testi\inst{2,3,4}\and
          M. Tazzari\inst{5}\and 
          A. Miotello\inst{2}\and
          P. Caselli\inst{1}\and
          M. Dunham\inst{6}\and \\
          I.W. Stephens\inst{7}\and
          T.L. Bourke\inst{8}
          }

   \institute{Max-Planck-Institute for Extraterrestrial Physics (MPE), Giessenbachstr. 1, D-85748 Garching, Germany\\
              \email{cagurto@mpe.mpg.de}
         \and
             ESO, Karl Schwarzschild str. 2, D–85748 Garching, Germany
         \and
         	 INAF-Osservatorio Astrofisico di Arcetri, L.go E. Fermi 5, Firenze, 			  50125, Italy
         \and 
              Excellence Cluster Universe, Boltzmannstr. 2, D-85748 Garching, Germany
         \and
             Institute of Astronomy, University of Cambridge, Madingley Road, CB3 0HA Cambridge, UK
         \and    
         	 Department of Physics, State University of New York at Fredonia, 280 Central Ave, Fredonia, NY 14063
         \and 
         	 Harvard-Smithsonian Center for Astrophysics, 60 Garden
Street, Cambridge, MA 02138, USA
		 \and 
         	 SKA Organisation, Jodrell Bank Observatory, Cheshire SK11 9DL, UK
             }

   \date{Received June 19, 2018; accepted January 14, 2019}

 
  \abstract
   {A good constraint of when the growth of dust grains from sub-micrometer to millimeter sizes occurs, is crucial for planet formation models. This provides the first step towards the production of pebbles and planetesimals in protoplanetary disks. Currently, it is well established that Class II objects have large dust grains. However, it is not clear when in the star formation process this grain growth occurs.}
   {We use multi-wavelength millimeter observations of a Class I protostar to obtain the spectral index of the observed flux densities $\alpha_\mathrm{mm}$ of the unresolved disk and the surrounding envelope. Our goal is to compare our observational results with visibility modeling at both wavelengths simultaneously.}
   {We present data from NOEMA at 2.7\,mm and SMA at 1.3\,mm of the Class I protostar, Per-emb-50. We model the dust emission with a variety of parametric and radiative transfer models to deduce the grain size from the observed emission spectral index.}
   {We find a spectral index in the envelope of Per-emb-50 of  $\alpha_{\rm env}$=$3.3\pm0.3$, similar to the typical ISM values. The radiative transfer modeling of the source confirms this value of $\alpha_{\rm env}$ with the presence of dust with a $a_\mathrm{max}$$\leq$100\,$\mu$m. Additionally, we explore the backwarming effect, where we find that the envelope structure affects the millimeter emission of the disk.}
   {Our results reveal grains with a maximum size no larger than $100$\,$\mu$m in the inner envelope of the Class I protostar Per-emb-50, providing an interesting case to test the universality of millimeter grain growth expected in these sources.}

\maketitle
\section{Introduction}

Disks and envelopes around protostars play a fundamental role in the process of planet formation since they contain the ingredients out of which planets are formed \citep{Testi2014}. \\
Thanks to detailed studies of protoplanetary disks at several sub-mm and mm wavelengths such as HL Tau \citep{Carrasco2016}, CY Tau, DoAr 25, and FT Tau \citep{Perez2015,Tazzari2016} , it is now well established that the radial profiles of their grain size distributions are compatible with millimeter size grains.
However, it is not yet clear at which stage of the star and planet formation process dust grains start to efficiently coagulate and evolve from $\mu$m size particles to macroscopic dimensions. 
\\
\cite{Ormel:2009dq} studied in detail the possibility of grain growth in pre-stellar cores and found that while it is easy to grow to micron size particles, the growth to millimeter or centimeter size pebbles requires high densities and relatively long timescales of $\sim$$10^{7}$ yr, much longer than the lifetimes of dense cores. This is also explored recently in \citet{Chacon2017}, where they calculate the grain size in the center of the pre-stellar core L1544, finding that only in the central 300\,AU, grain size can grow to about $200$$\mu$m.
\\ 
In the earliest protostellar phases, e.g. during the Class 0 stage, the protostar is fully embedded in the parent envelope, while in the Class I phase, the envelope is partially dissipated and the disk emission can be better separated from the envelope. Therefore, Class I protostars can more easily address the start of planetesimal formation and constrain the initial conditions of the evolution of protoplanetary disks. \\
The possibility for the first large solids to assemble during the early phases of disk evolution would have important implications. If the process starts already in the Class I stage it would imply a much more effective and rapid planetesimal formation phase in the disks. In fact, if large (mm to cm-size) dust particles from the inner envelope \citep{Chiang12, Tobin2013, Miotello14} are deposited in the disk at large radii during the disk formation stage, they would be much less affected by the radial transport and fragmentation processes, which adversely affect the growth from sub-micron particles, and large dust aggregates could form \citep{Birnstiel2010}. 

The advantage of studying protostars at millimeter wavelengths is that the dust emission from the envelope and the disk is mostly optically thin. In this wavelength range, the dust opacity coefficient $\kappa_\mathrm{\nu}$, can be approximated by a power law $\kappa_\mathrm{\nu} \propto \nu^{\beta_\mathrm{mm}}$, where $\beta_\mathrm{mm}$ is the millimeter dust opacity spectral index, and is directly related to the maximum size of the grain \citep{Natta07}. In the presence of very large grains, much larger than the observing wavelength, the opacity becomes gray (only the geometrical cross section of the grains is relevant) and $\beta_\mathrm{mm}$= \ 0. Values of $\beta_\mathrm{mm}$ can be estimated by measuring the slope $\alpha_\mathrm{mm}$ of the sub-mm spectral energy distribution (SED), $F_\mathrm{\nu} \propto \nu^{\alpha_\mathrm{mm}}$. When the Rayleigh--Jeans approximation is applicable, the spectral index of the observed flux densities, $\alpha_\mathrm{mm}$, would translate to a power law index of the dust opacity $\beta_\mathrm{mm} = \alpha_\mathrm{mm}$ - 2. While values around $\alpha_\mathrm{mm}\sim$3.7 represent size distribution similar to interstellar medium (ISM) particles \citep{Natta07,Testi2014}. Classical protoplanetary disks around Class II objects, present clear signs of dust coagulation, with $\alpha_\mathrm{mm}$$\leq3$ (\citealt{Testi2014}, and references therein). \\
Previous observations of Class 0 protostars by \citet{Chiang12}, \citet{Jorgensen07} and \citet{Kwon09}, indicate spectral indexes $\alpha_\mathrm{mm} \sim 3$, which is shallower than the ISM, but not quite as steep as Class II disks. However, Class 0 objects, are affected by the presence of powerful accretion of material from the envelope and jets, e.g. \citep{Tobin2013}, making them difficult to observe and model. In contrast, Class I protostars have less massive envelopes, which provides a more cleaner analysis of the dust properties since the envelope and disk emission can be separated. \\
Here we present a dual wavelength analysis and modeling on the Class I protostar Per-emb-50, in the Perseus star forming region. Observations and data reduction are described in Sec. 2. In Sec. 3 we present our observational analysis. The modeling and discussion are presented in Section 4 and 5 respectively. Conclusions and future work are in Section 6.
\begin{figure*}[h]
\centering
\includegraphics[width=0.7\textwidth]{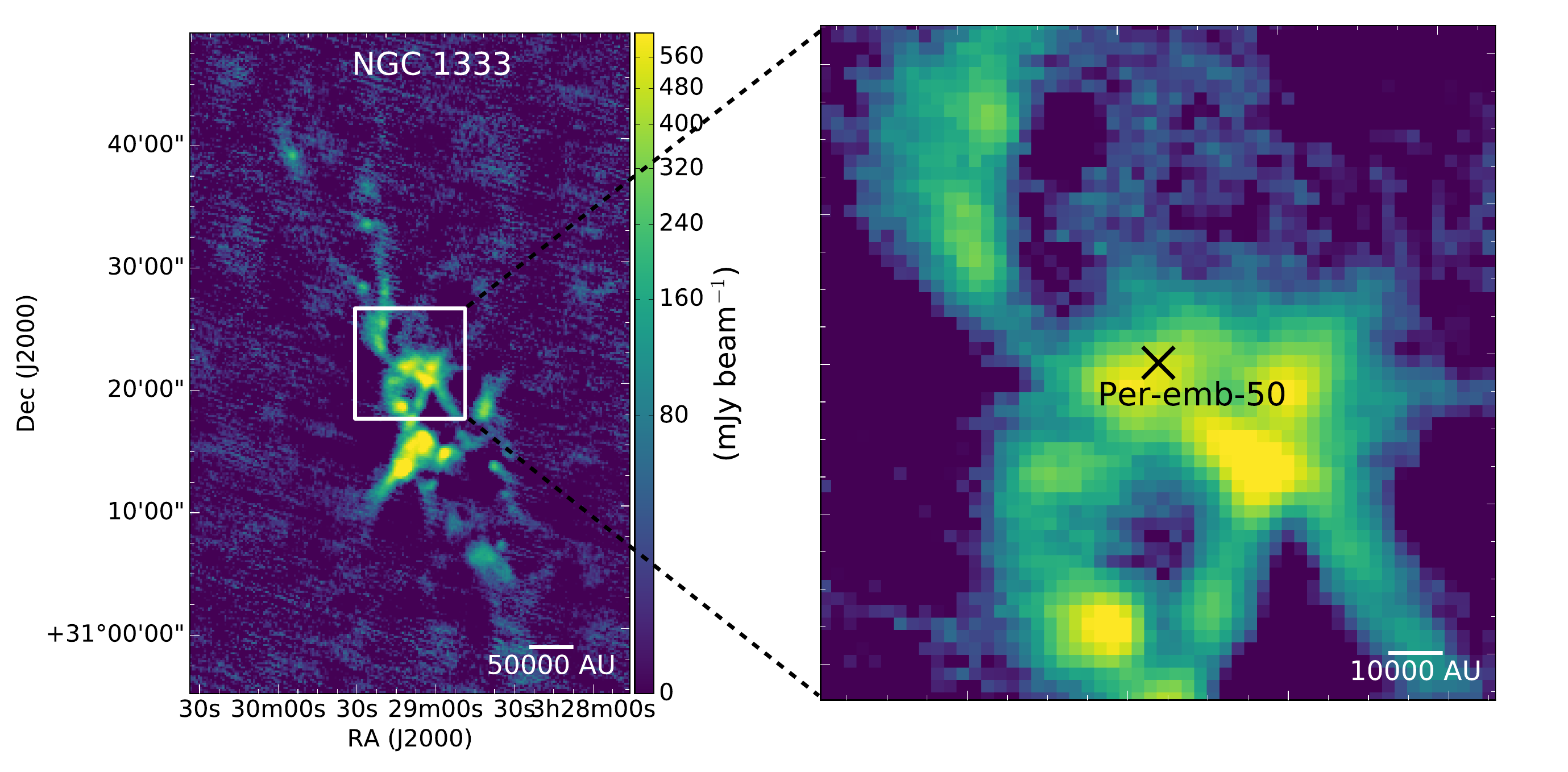}
\caption{(Left) Continuum map of the NGC1333 complex at 1.1 mm wavelength. (Right) Zoom-in to the direct enviroment of Per-emb-50. The map is adapted from the Bolocam survey at the Caltech Submillimeter Observatory (CSO) by \citet{Enoch2006}.}
\label{fig-ngc}
\end{figure*}
\section{Observations and Data reduction}
\subsection{The source}
Per-emb-50 is a protostar located in the active cluster forming region NGC1333 in the Perseus cloud (see Figure \ref{fig-ngc}), at a recently revised distance of 293~pc \citep{Ortiz-Leon18,Zucker2018}.
It is classified as a Class I protostar from the slope of its SED in the near-, mid-infrared ("Cores to Disks" or c2d Spitzer Legacy project from, \citealt{Evans2003}). Based on Bolocam 1.1mm data, the bolometric temperature is T$_\mathrm{bol}$=254$\pm$23 K. The rescaled bolometric luminosity is L$_\mathrm{bol}$=13.7$\pm$3 L$_{\odot}$, making it one of the brightest Class I sources in Perseus.
\\
High angular resolution observations conducted at 8mm in the VLA Nascent Disk and Multiplicity (VANDAM) survey, provide a lower limit for the disk mass and outer disk radius. The rescaled values from \citet{SeguraCox2016} for mass and radius are: M$_\mathrm{disk}$=0.28--0.58 M$_{\odot}$ and r$_\mathrm{out}$=27--32 AU, respectively. \\
Literature values for envelope mass, disk mass, disk inclination, and other parameters are presented in Table \ref{literature}.
We note that some of these physical parameters were calculated using the 230 pc from \citet{Hirota2008} or 250 pc in the case of Bolocam observations, therefore, we rescale the limits taking into account the different distance adopted.\\
Even though Per-emb-50 presents a small disk at 8\,mm, it is the perfect candidate for studying the growth in the inner envelope and their dust properties.
\begin{table}[ht]
\caption{Parameters from literature.}             
\label{literature}      
\centering                          
\begin{tabular}{c c c c}        
\hline\hline           
Source & Per-emb-50 & New value & Ref  \\    
\hline\hline                        
     RA$_\mathrm{J2000}$  & 03:29:07.76 & -- & 1 \\      
     Dec$_\mathrm{J2000}$ & +31:21:57.2 & -- & 1  \\
     $\mathrm{L_{bol}}$ $(L_{\odot})$& $10\pm$3.0& 13.7$\pm$3.0 &1\\
     $\mathrm{M_{env}}$ $(M_{\odot})$ & $1.62\pm$0.16  & 2.2$\pm$0.16 & 1  \\
     $\mathrm{PA}$ (deg) & $170\pm$0.3     & -- & 2\\
     $\mathit{i}$ (deg)* & $67\pm$10     & -- & 2\\
     $\mathrm{M_{disk}}$ $(M_{\odot})$ & 0.18 -- 0.36 & 0.28 -- 0.58  & 2  \\
     $\mathrm{R_{disk}}$ (au) & 21.9 -- 25.7 & 27.3 -- 32.1   & 2  \\
     F$_{1.1\mathrm{mm}}$ (mJy)& $612\pm18$ & -- & 3 \\
\hline                                   
\end{tabular}
\tablebib{(1) \citet{Enoch09}; (2) \citet{SeguraCox2016}; (3, single dish observation) \citet{Enoch2006}}
\tablefoot{
\tablefoottext{*}{$\mathit{i}=0$ is a face-on disk.}
}
\end{table}
\subsection{SMA observations}
The Submillimeter Array (SMA) data shown in this paper are from the MASSES legacy program (Mass Assembly of Stellar Systems and their Evolution with the SMA, PI: I.W. Stephens, M. Dunham; e.g.,\citet{Stephens2018}). \\ 
Per-emb-50 was observed at 
1.3~mm with the receiver centered at 220.69 \mbox{GHz}, in the Extended (eight antennas) and Subcompact (seven antennas) configuration with ASIC (Application Specific Integrated Circuit) correlator during September 2015 and November 2014, respectively. Additionally, Per-emb-50 was observed during October 2015 with SWARM (SMA Wideband Astronomical ROACH2 Machine) correlator at 1.3mm in extended configuration (see Table \ref{summary} for more details). Weather conditions were good, with zenith optical depths at 220\,GHz of $\tau_{220}$ = 0.07 -- 0.15. \\
Calibration was done in MIR while imaging was done in MIRIAD \citep{Sault1995}, using the standard calibration procedure. We inspected the amplitudes and phases of the calibrators on each baseline in order to look for variations or noisy data, which were manually flagged. Corrections for system temperatures were applied in order to calibrate the atmosphere attenuation in the visibility amplitudes. Detailed information of the calibration can be found in \citet{Stephens2018}. \\
The quasars 3C454.3 and 3C84 were used as bandpass and phase calibrators. The absolute flux was calibrated on Uranus, with $\sim$20\% of flux calibration uncertainty. For the purpose of this work, we use the 1.3~mm data in the Subcompact and Extended array configurations, with projected baselines in the range of 23-119 k$\lambda$. The resulted combined beam was 1.$^{\prime\prime}$72 $\times$ 1.$^{\prime\prime}$40 at P.A. 50.80$^{\circ}$. \\

\subsection{NOEMA observations}

The 2.7~mm data presented in this work were obtained with NOEMA, the IRAM\footnote{http://www.iram-institute.org/} NOrthern Extended Millimeter Array. 
The observations were performed on November 6th and 12th, 2016. The array was in the C compact configuration, with 8 antennas (8C) in operation during the first track, and 6 antennas (6C) in operation for the second.
Antennas were based on stations E10, W20, W10, N20*, E18, N11*\footnote{Stations with * correspond to antennas not available for the second day track}, N17 and E04.  The projected baselines were from 7.8 k$\lambda$ and 102 k$\lambda$. \\
Per-emb-50 was observed for hour angles from -5.8 to 1.5 h for 8C, and from -5.3 to 1.4 h for 6C. In total we spent 9 hours on source.
0333+321 was used as phase/amplitude calibrator. The sources LkHa101 and MWC349 were used for the flux calibration, while the quasars 3C84 and 3C454.3 were used for the bandpass calibration. 
We consider an absolute flux uncertainty of 10\%. The total bandpass for the 110 GHz continuum measurement was 2 GHz. 
Data reduction and image synthesis were carried out using the GILDAS software \citep{Gildas2000} with the procedure of MAPPING> Selfcal. The continuum map (Fig. 2) was produced using natural weighting and the resulting beam size is 2.$^{\prime\prime}$1 $\times$ 1.$^{\prime\prime}$6 at P.A. 34.84$^{\circ}$. The clean map has a $rms$ noise level of 2.1 mJy beam$^{-1}$.
\begin{figure}
\centering
\includegraphics[width=9cm]{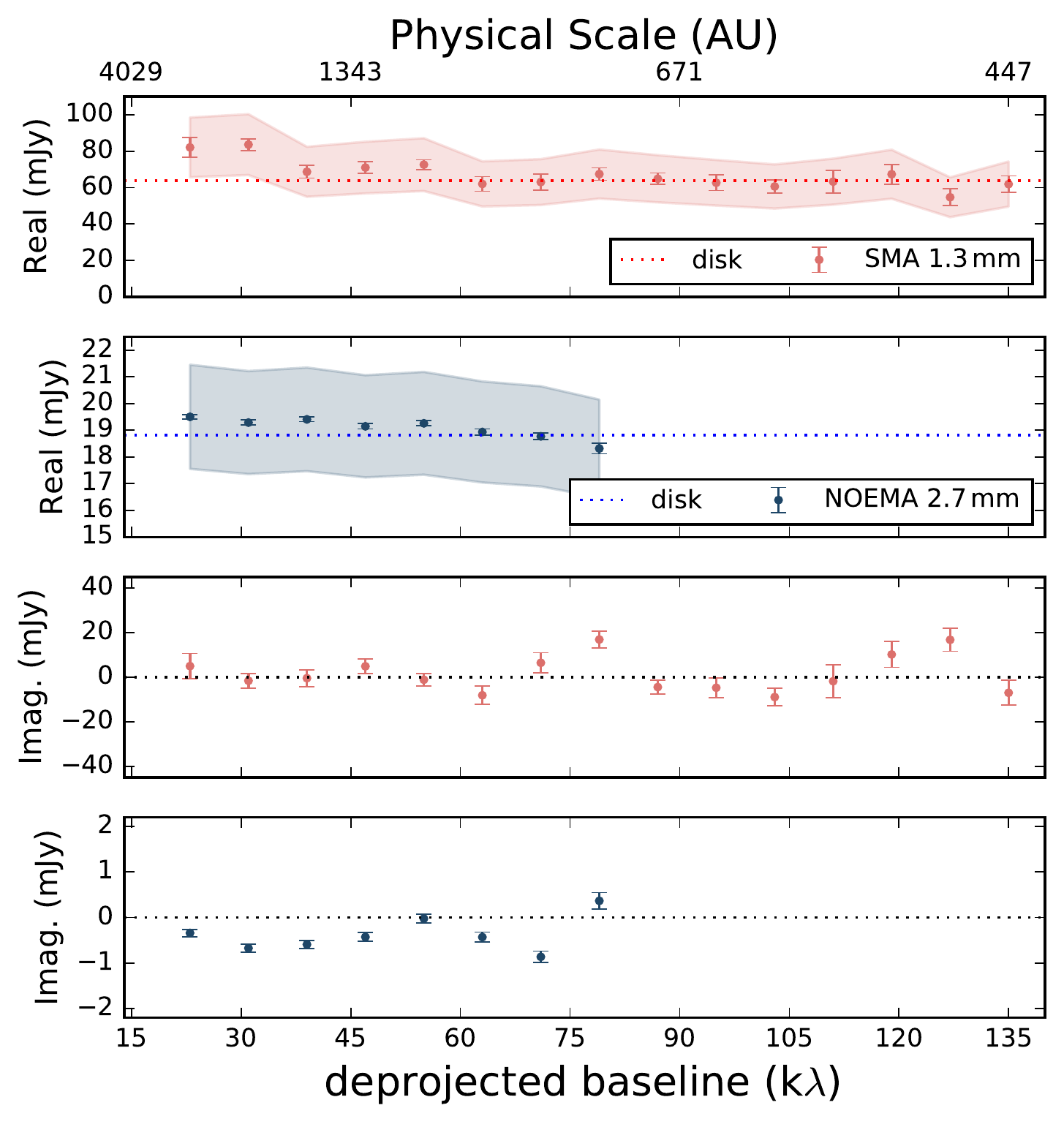}
\caption{Real and imaginary parts of the measured visibilities of Per-emb-50 as a function of the deprojected baseline, assuming the PA and $i$ from Table 1. The data is averaged in 8 k$\lambda$ bins. The error bars in the real parts show the statistical standard errors of visibilities in each bin. Red and blue shaded areas show the 20 per cent and 10 per cent flux calibration uncertainties of the SMA and NOEMA data, respectively. Red and blue dashed lines are the disk average fluxes using baselines larger than 47 k$\lambda$.}
\label{visibilities}
\end{figure}

\begin{figure}
\centering
\includegraphics[width=9cm]{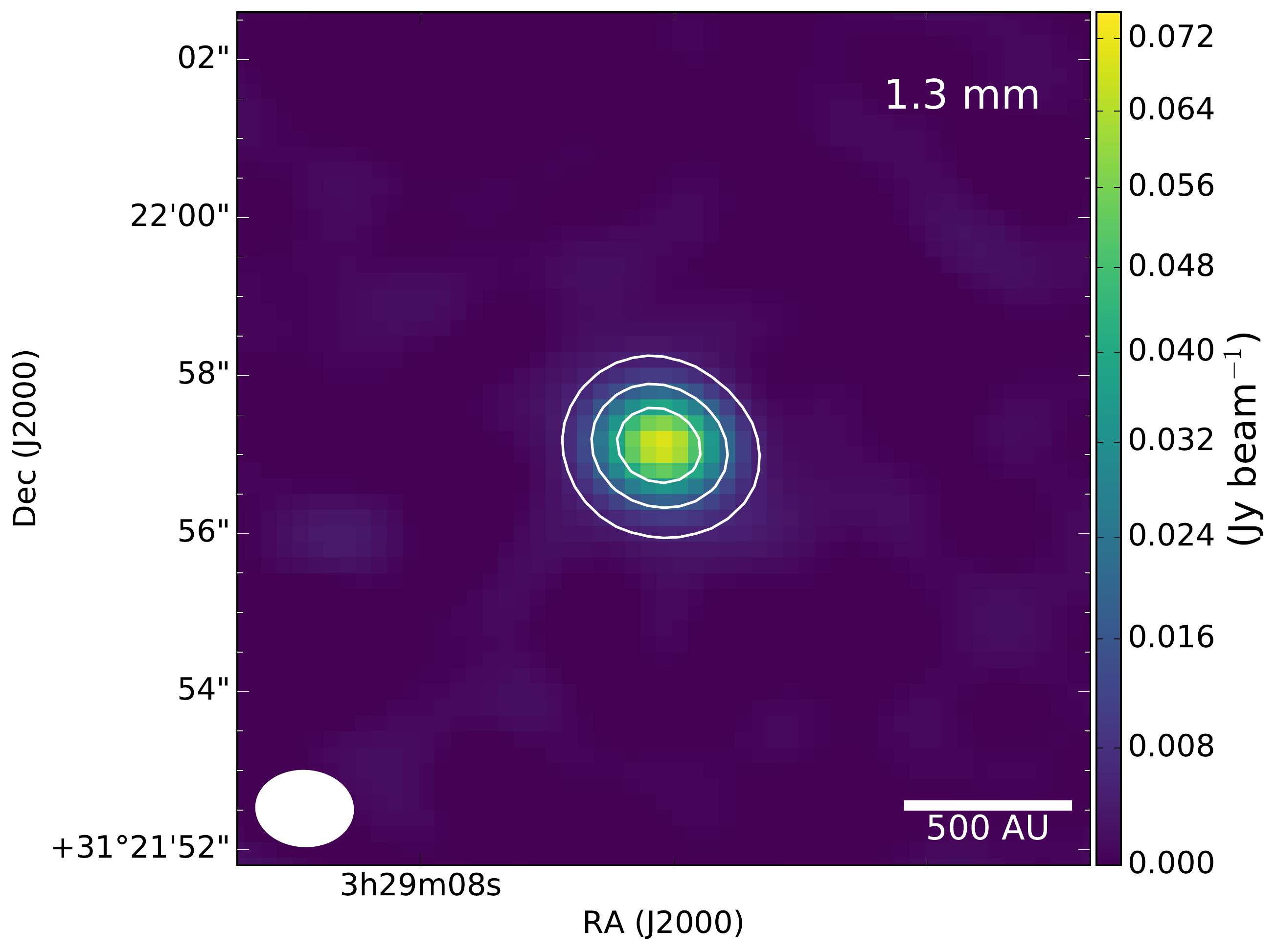}
\includegraphics[width=9cm]{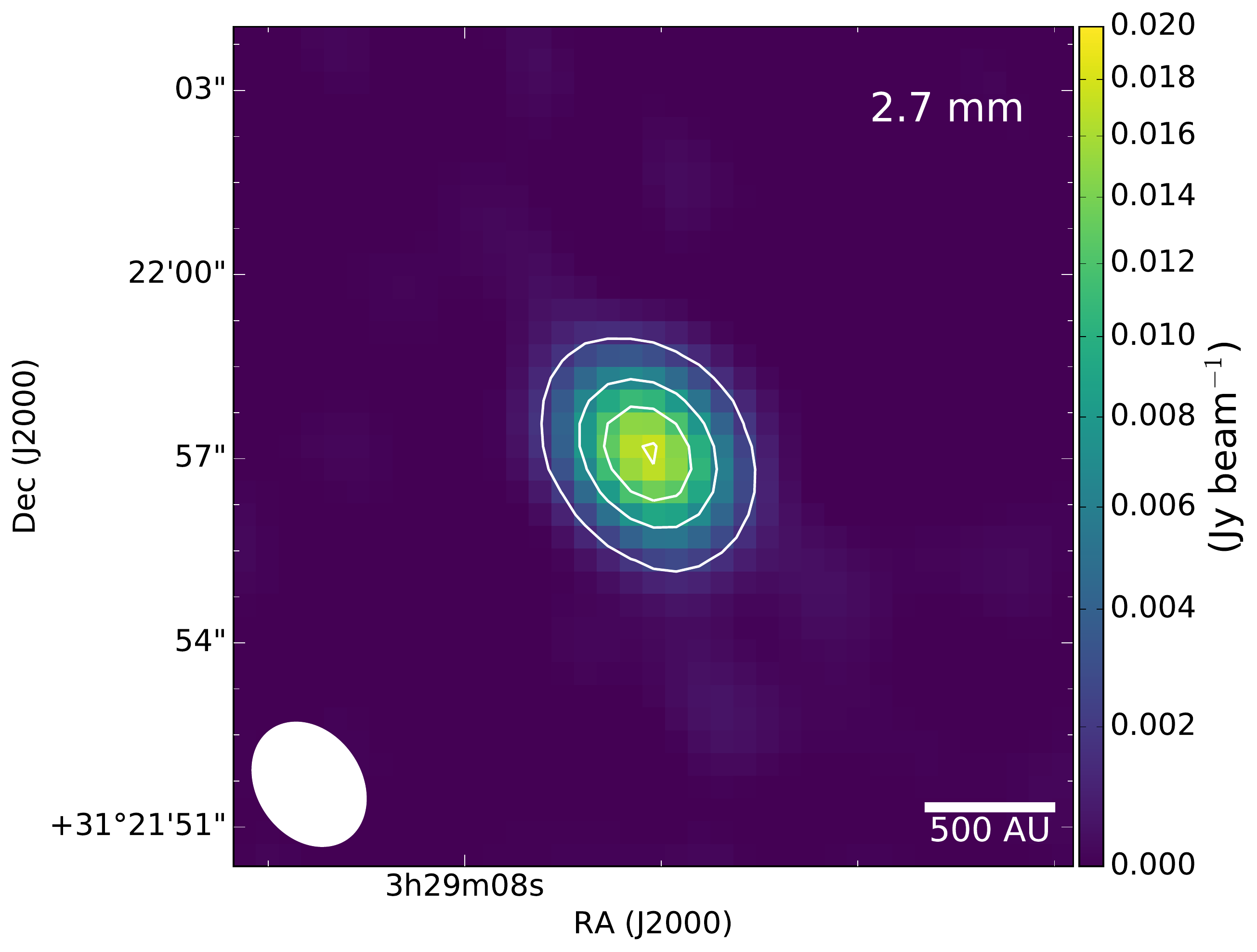}
\caption{Continuum map of Per-emb-50 at 1.3\,mm (SMA) and 2.7\,mm (NOEMA) wavelengths. The synthesized beam FWHM is represented as a white ellipse in the bottom-left corner of each map. For SMA and NOEMA data, the contours start at 76 mJy beam$^{-1}$ and 20 mJy beam$^{-1}$, respectively, and both increase in 25\% intervals. }
\label{maps}
\end{figure}

\begin{table*}
\caption{Summary of Observations}             
\label{summary}      
\centering          
\begin{tabular}{c c c c c l l l l}     
\hline      
\hline                      
Observatory & Representative & Date & Flux  & Bandpass  & Flux & Array &  Synthesized & P.A.\\
 & Frequency &  & Calibrator  & Calibrator  &  & Configuration &  Beam  & \\
& (GHz) &  & &  & (Jy) &  &($^{\prime\prime}$) &($\hbox{$^\circ$}$)\\
\hline
SMA & 220 & Nov 27 2014 & Uranus  & 3c84 & 11.64 & Subcompact & 1.2$\times$0.96 & 86.9
 \\
&  &  &   & 3c454.3 & 16.71 & Subcompact \\
& & Sep 15 2015 & Uranus & 3c84 & 11.64 & Extended \\
& & Oct 29 2016 & Neptune & 3c84 & 13.8 & Extended \\
& &  &  & 3c454.3 &  &  \\
\hline
NOEMA & 109\tablefootmark{} & Nov 6 2016 & 0333+321    & 3c454.3& 14.03& 8C & 2.2$\times$1.7 & 35 \\
   & & Nov 12 2016 & 0333+321    & 3c84 & 24.80 & 6C  \\
\hline                  
\end{tabular}
\end{table*}
\section{Observational Analysis}
Since we are working with interferometric data, the best way to analyze our source is working on the visibility domain. This is to avoid biases in the model--data comparison that are introduced by the CLEAN algorithm, u--v sampling, and the imaging process.\\
In Fig.\ref{visibilities} we plot the real visibility as a function of the deprojected baseline length ($\mathrm{uv}$-distance). The deprojected uv-distances are given by $R$=$\sqrt{\smash[b]{d_{a}^{2} + d_{b}^{2}}}$, where $d_{a}$=$\sqrt{u^{2}+v^{2}}\sin \phi$ and $d_{b}$=$\sqrt{u^{2}+v^{2}}\cos\phi \cos i$, $\phi$=$\arctan(v/u)-PA$ \citep{Lay1997}. The values for inclination $i$, and position angle, \rm{PA}, are presented in Table\ref{literature}. \\
In Fig.\ref{maps}, we show images of the SMA and NOEMA observations. Per-emb-50 appears as a point source in these two images, so we do not resolve the embedded disk. Consequently, since the disk is unresolved, then it contributes as a constant component at all baselines. At long baselines, we expect that the amplitude of the visibility is dominated by the disk component, while in shorter baselines the resolved envelope dominates. \\
For Per-emb-50, the value of the amplitudes start becoming constant above 47 k$\lambda$ (see Fig.\ref{visibilities}) for both wavelengths. We assume that the emission from those baselines belongs to the embedded disk, where the average values at 2.7 mm and 1.3 mm are: $F_\mathrm{disk}^\mathrm{2.7mm}$=$18.82\pm 0.13$ mJy and $F_\mathrm{disk}^\mathrm{1.3mm}$=$63.85\pm 4.2$ mJy.

The spectral index $\alpha_\mathrm{mm}$ can be calculated through the flux ratio between two wavelengths,
\begin{equation}
\alpha_\mathrm{mm} = \frac{\ln \mathrm{F_{1}}-\ln \mathrm{F}_{2}}{\ln \mathrm{\nu}_{1}-\ln \mathrm{\nu}_{2}} \,
\end{equation}
Using the fluxes between the u--v ranges 47--80~k$\lambda$ at 1.3 and 2.7\,mm, we obtain therefore the average value $\alpha_\mathrm{mm}$ in the unresolved disk, which is $\alpha_\mathrm{disk}$=1.71$\pm$0.3. \\
As shown in the Fig.\ref{visibilities}, an excess of emission is present at short baselines ($<$47 k$\lambda$) at 2.7\,mm and 1.3\,mm, which correspond to physical scales of 1500--3000~AU. The excess values at these baselines, after subtracting the disk visibilities, are: $F_\mathrm{ex}^\mathrm{2.7mm}$=$0.52\pm0.1$~mJy and $F_\mathrm{ex}^\mathrm{1.3mm}$=$10.1\pm5$~mJy. The excess, even if not very pronounced at 2.7\,mm, is detected at 1.3\,mm, therefore this indicates the presence of extended emission related to the inner envelope.
Considering the average fluxes at these very short baselines, and using the same u--v distances ranges at both wavelengths, we recover an average value for $\alpha_\mathrm{env}$ bigger than the typical ISM values, $\alpha_\mathrm{env}$=$4.0\pm0.8$. 
\begin{figure}
\centering
\includegraphics[width=6cm]{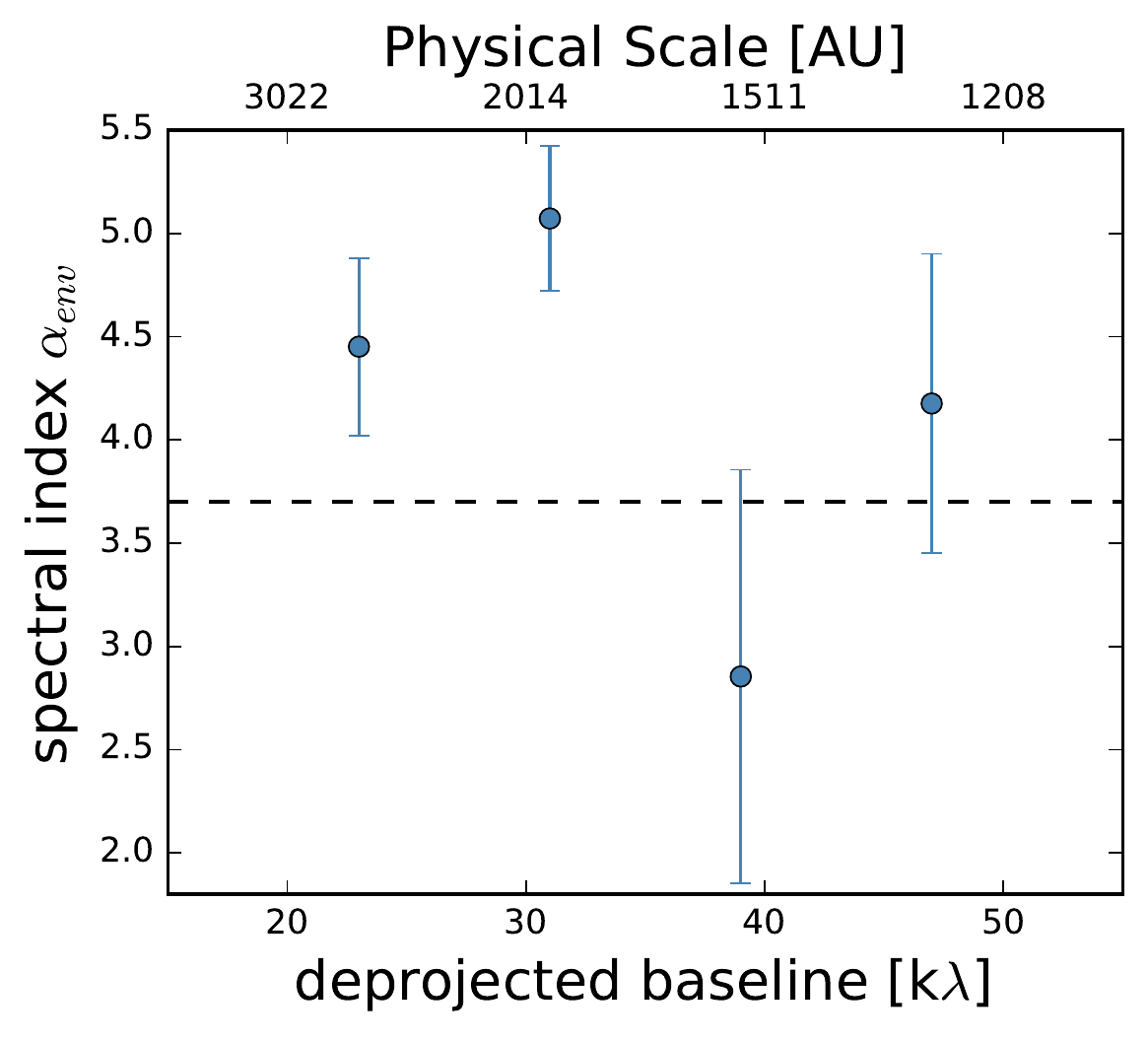}
\caption{Spectral index of the envelope as a function of deprojected baseline. The black dashed line represents the typical value of $\alpha$$\sim$3.7 related to grain properties in the diffuse interstellar medium.}
\label{alpha_env}
\end{figure}
The uncertainty in $\alpha_\mathrm{disk}$ and $\alpha_\mathrm{env}$ are estimated following the procedure shown in Appendix A, with the absolute flux uncertainty of 10\% for 2.7~mm data and 20\% for 1.3~mm data added in quadrature. We present $\alpha_\mathrm{env}$ and its change as a function of deprojected baseline in Fig.\ref{alpha_env}. 
If we translate this value to the spectral index of the dust opacity, we obtain a $\beta_\mathrm{env}$$\sim$2.0, which is similar to the values found in the ISM.
\\
\\
These preliminary results are showing a discrepancy with previous studies on spectral indexes in Class I protostar or even younger sources with dust opacity indexes $\alpha_\mathrm{mm}$<3 \citep{Miotello14,Kwon09,Chiang12}. To investigate possible explanations, we will perform a partial and full radiative transfer modeling on envelope and disk to take into account possible deviations from the optically thin and Rayleigh Jeans regimes, which can affect the values of $\alpha_\mathrm{mm}$.
\section{Modeling}
In order to model the Class I protostar and compare with the observations, we consider appropriate physical structure and conditions of the source, including the envelope structure, density, properties of the dust grains, and we predict the 1.3 and 2.7 mm emission with a u--v modeling described below. \\
In the first step, we fit the Per-emb-50 data with a parametric modeling in uv space (Section 4.1) in order to address the $\alpha$ values, as well
as visibility comparisons. Afterward, we use the radiative transfer tool RADMC-3D \citep{Dullemond2012} in two ways: (a) to apply the modeling approach of \citet{Miotello14}, where the disk and envelope are modeled separately (Section 4.2), (b) to compute the emission for the new modeling presented in this work, that include a self-consistent radiative transfer model for the disk and the envelope (Section 4.3). 
In the following sections, we discuss and compare the details of the results of each modeling case.
\subsection{Parametric Model}
We implement a model that consists of an extended envelope described by a Gaussian, and an unresolved disk (point source) that has constant flux at all baselines.
Therefore, the combined amplitude profile, that depends on the $\mathrm{uv}$ distance, defined as $\mathrm{\sqrt{\mathrm{u}^{2}+\mathrm{v}^{2}}}$, and frequency, $\nu$, is described by:

\begin{equation}
f($uv$,\mathrm{\nu}) = F_{e}\left(\frac{\mathrm{\nu}}{\nu_\mathrm{1.3mm}}\right)^{ \alpha_{e}} \mathrm{exp} \left ({\frac{-{(\mathrm{uv})}^{2}}{2\sigma^{2}}} \right ) + F_{d}\left(\frac{\nu}{\nu_\mathrm{1.3mm}}\right)^{\alpha_{d}},
\end{equation}
where $\mathit{F_\mathrm{e}}$ and $\mathit{F_\mathrm{d}}$ are the flux density from the Gaussian emission (extended envelope) and point source emission (unresolved disk) respectively, $\alpha_\mathrm{e}$ and $\alpha_\mathrm{d}$ are the spectral indexes of the two components, and $\sigma$ is the width of the Gaussian given by $\sigma\approx \mathrm{FWHM}/2.355$. \\
In this simple model, we first set the flux from the disk at 1.3~mm based on the average value reported in Sec. 3, $\mathit{F_\mathrm{d}}$=63 mJy. Then, four parameters are explored: $F_\mathrm{e}$, $\alpha_\mathrm{e}$, $\alpha_\mathrm{d}$ and $\sigma$. The Markov chain Monte Carlo (MCMC) method, implemented as a python package $emcee$ \citep{emcee}, is utilized to calculate the posterior probability distributions of each of these parameters. For each model we used the 750 steps after the burn-in and 400 walkers (see Appendix B for more details). The results from this simple model will be discussed in the following section.
\subsubsection{Parametric model results}
\begin{figure}[h!]
\centering
\includegraphics[width=7cm]{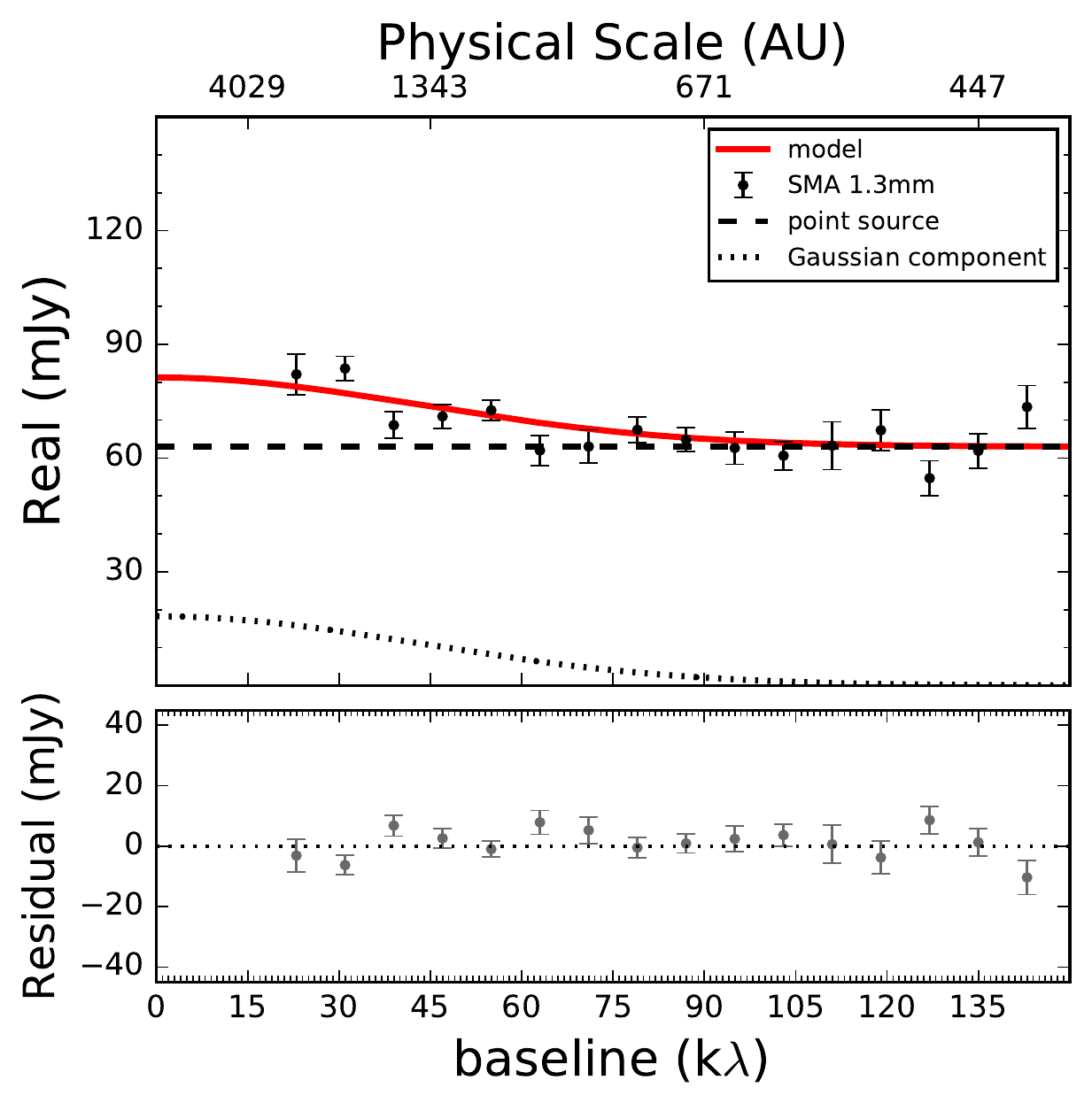}
\includegraphics[width=7cm]{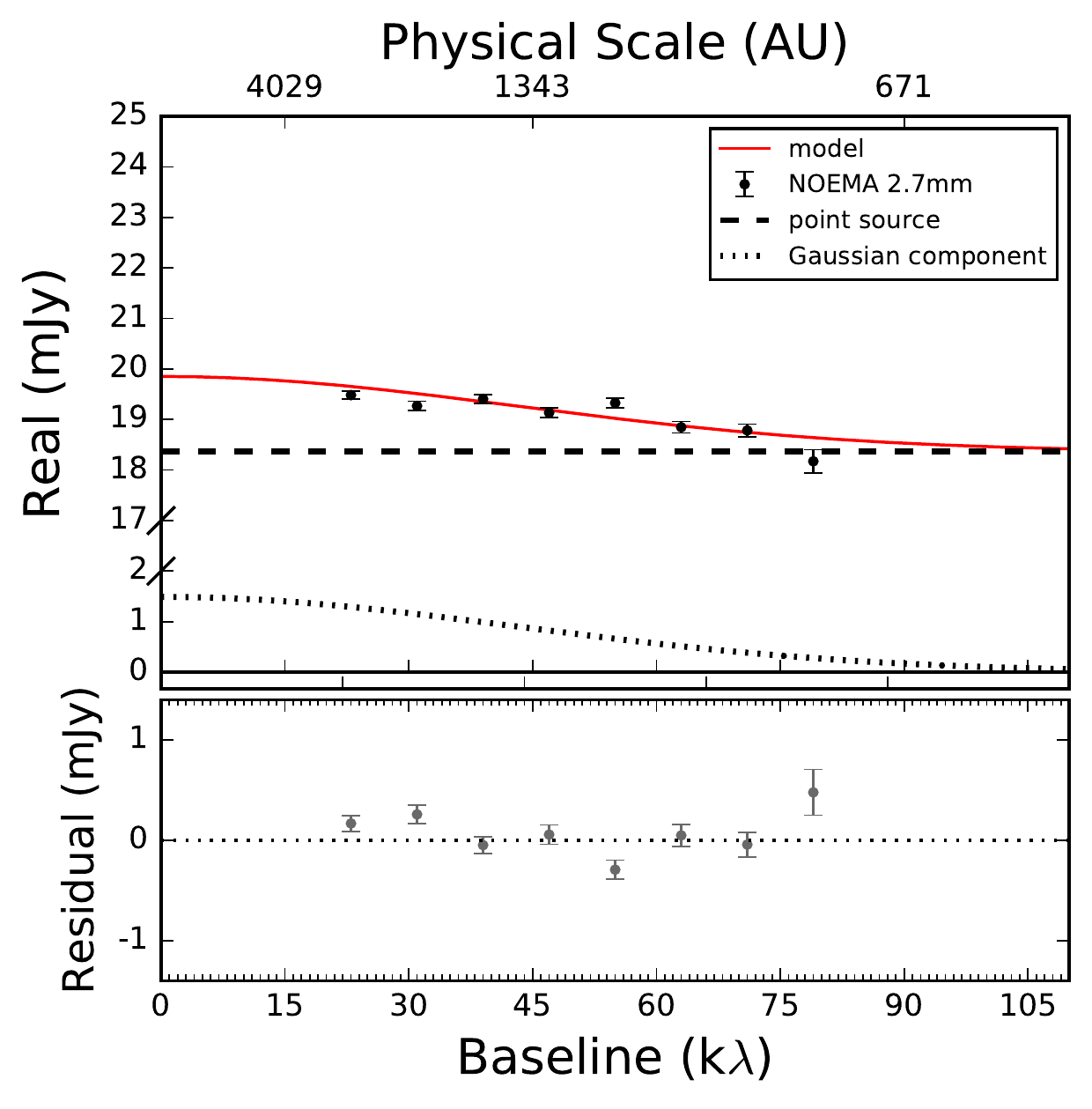}
\caption{Black points are the real part of the visibilities as a function of the baseline length. Red curves show the best fit model, while the dashed and dotted lines indicate its point source and Gaussian components, respectively. Bottom panels show the residual between the model and data.}
\label{parametric}
\end{figure}

In  Fig. \ref{parametric}, model visibilities are compared with observational data at each u$-$v sample and wavelength. In Table \ref{parametric-table} we present the best-fit parameters found for this parametric model. The values of the flux spectral index in the disk and envelope are consistent with the observational analysis (see Section 3), but their errors are highly dominated by the systematic error of absolute fluxes and the statistical error of the data. We estimate that the uncertainty on $\alpha_\mathrm{mm}$ for the envelope and disk using a simplistic approximation for non correlated errors is $\pm0.3$.\\ 
Additionally, from this simple model we can constrain the size of the region where the envelope emission arises, 
which the 1-sigma width (from Table \ref{parametric-table}) 43 k$\lambda$ (1405 AU). From the model we can derive the flux from the disk at 2.7\,$\mathrm{mm}$ and the prediction of the total flux at baseline=0 k$\lambda$ or zero spacing flux, $\mathit{F^{2.7\mathrm{mm}}_\mathrm{zero}}$ and $\mathit{F^{1.3\mathrm{mm}}_\mathrm{zero}}$. The results are shown in Table \ref{parametric-derived}.

We use the derived parameters from our parametric model (Table \ref{parametric-derived}) to estimate the zero spacing flux at 1.1\,mm, $\mathit{F_\mathrm{zero}^\mathtt{1.1mm}}$. We find the flux is only 127.4 mJy\,beam$^{-1}$ which is much lower than the single dish flux of 612$\pm$18 mJy reported by \citet{Enoch2006}. This discrepancy is related to the resolution of the observations. While our interferometric data is sensitive to the inner envelope of this source, the 31$^{\prime\prime}$ beam size of Bolocam is recovering the extended emission, which is affected by blending effects, especially in a crowded region such as NGC 1333 (see Fig. 1).
\begin{table}
\caption{Best Parametric Model}             
\label{parametric-table}      
\centering                          
\begin{tabular}{c c c c}        
\hline\hline           
Fit parameters &  &  \\    
\hline\hline                        
   $\mathit{F_\mathrm{e}}$ (mJy) & $17.0\pm$1.1 \\
     $\mathit{\alpha_\mathrm{e}}$ & $3.3\pm$0.3 \\
     $\mathit{\alpha_\mathrm{d}}$ & $1.7\pm$0.3 \\
     $\mathit{\sigma}$ (k$\lambda$) & $43.1\pm$0.5 \\ 
\hline                                   
\end{tabular}
\end{table}

\begin{table}
 \begin{center}
   \caption{Derived parameters from Parametric model}    
   \label{parametric-derived}
   \begin{tabular}{c c}        
   \hline\hline                 
      $\mathit{F_\mathrm{d}^{2.7\mathrm{mm}}}$ [mJy] & $18.4\pm0.7$ \\
      & \\
      $\mathit{F^{2.7\mathrm{mm}}_\mathrm{zero}}$ [mJy] & $19.8\pm2.2$ \\
      $\mathit{F^{1.3\mathrm{mm}}_\mathrm{zero}}$ [mJy] & $81.3\pm2$ \\       
      \\
      $\mathit{F^{1.1\mathrm{mm}}_\mathrm{zero}}$ [mJy] & $127.4\pm2$ \\

   \hline 
   \end{tabular}
 \end{center}
\end{table}

Since this simple model is not taking into account properties of the dust grains and density profiles for both the envelope and disk, we also analyze Per-emb-50 with more detailed dust radiative transfer models. 
\begin{table*}[h!]
\caption{Two--step model grid parameters}  
\label{grid2step}
\centering          
\begin{tabular}{c c c c}     
\hline\hline      
Parameter & Description & Values & Parameter Use \\
\hline
Stellar model parameters \\
d (pc) & Distance & 293 & fixed \\
$L_{\star}$ ($L_{\odot}$) & Photosphere luminosity & 13.7 & fixed \\
$T_{\star}$ (K) & Effective temperature & 5\,011 & fixed\\
$R_{\star}$ (AU) &  Stellar radius & 0.025 & fixed\\
\vspace*{2mm}
$M_{\star}$ ($M_{\odot}$)& Stellar mass & 2.9 & fixed\\
\hline
Disk model parameters & & & \\
$R_\mathrm{in}$ (AU) & Disk inner radius & 0.1 & fixed  \\
$R_\mathrm{out}$ (AU) & Disk outer radius & 25 27 30 32 34 36 & varied  \\
$M_\mathrm{disk}$ ($M_{\odot}$) & Disk mass & 0.05 0.1 0.15 0.2 0.3 0.4 & varied \\
$\Sigma_\mathrm{disk}$ (gr $\mathrm{cm^{-2}}$) & Disk surface density & 54.7--908 & varied\\
\vspace*{2mm}
$a_\mathrm{max}^{disk}$ ($\mu$m) & Disk maximum grain size & 500 1\,000 5\,000 10\,000 20\,000 & varied \\
\hline
RADMC-3D / Envelope parameters \\
$r_\mathrm{in}$ (AU)&  Envelope inner radius  & 25 27 30 32 34 36 & varied \\
$r_\mathrm{out}$ (AU) & Envelope outer radius & 8\,800 & fixed \\
$R_\mathrm{rot}$ (AU) & Centrifugal radius & 100--1000 & varied \\
$\rho_{0}$ (gr $\mathrm{cm^{-3}}$) &  Density in the equatorial plane at $R_\mathrm{rot}$ & 0.5$\times10^{-20}$--20.0$\times10^{-20}$ & varied \\
\vspace*{2mm}
$a_\mathrm{max}^{env}$ ($\mu$m) & Envelope maximum grain size & 0.1--1\,000 & varied \\
\hline
\end{tabular}
\tablefoot{Each model is calculated with 1$\times10^{6}$ photons for the thermal Monte Carlo.}
\end{table*}
\subsection{Two--step Model} 
For this model, we adopted the procedure described by \citet{Miotello14}, where they analyzed two Class I protostars with a 2$-step$ model. The disk is modeled adopting the two-layer model by \citet{Dullemond2001}, whose output spectrum is taken as central source of illumination in the envelope model. The envelope, on the other hand, is modeled using RADMC-3D \citep{Dullemond2012}.
\subsubsection{Modeling Protostar and Disk}
We adopt a simple disk model heated by protostellar radiation. We calculate the properties of the central protostar, assuming that it emits black body radiation, characterized by a radius $R_{\star}$, effective temperature $T_\mathrm{eff}$, and mass $M_{\star}$. 
To obtain $T_\mathrm{eff}$ we assume that Per-emb-50 lies along the birthline for intermediate mass stars by \citet{Palla1990}. Given the rescaled bolometric luminosity $L_{\rm bol}$ reported in Table \ref{literature}, we estimate $T_{\rm eff}$=5011 K. With $L_{\rm bol}$ and $T_{\rm eff}$, we can estimate $R_{\rm eff}$ using the Stefan-Boltzmann law:
   \begin{equation}
      R_{\rm eff} = \left( \frac{L_{\rm bol}}{4\pi \sigma T_{\rm eff}^4} \right)^{1/2} \,,
   \end{equation}
Then, with $R_{\rm eff}$=5.01 $R_{\odot}$, we use the mass vs. radius relation for a spherical protostar accreting at a rate of $10^{-5}$ $M_{\odot}$ yr$^{-1}$ from \citet{Palla1991}, to deduce an effective mass of $M_{\rm eff}$=2.9 $M_{\odot}$ (Table. \ref{grid2step}).
Additionally, we add a disk structure defined by an inner and outer radius, $r_\mathrm{in}$ and $r_\mathrm{out}$, an inclination angle $i$, and a dust surface density profile that follows a simple power law,
   \begin{equation}
      \Sigma(R) = \Sigma_{0}\left( \frac{r_{out}}{r_{\Sigma_{0}}} \right)^{-p} \,,
   \end{equation}
where $\Sigma_{0}$ is the surface dust density fixed at r$_{\Sigma_{0}}$= 1 AU from the central protostar, and where p=1 since the quality of the data is not sensitive enough to discriminate between different values of p. The disk inclination $i$ is fixed to 67$^{\circ}$ as found by \citet{SeguraCox2016}. Since the mm-SED is not sensitive to $r_\mathrm{in}$, we set $r_\mathrm{in}$=0.1 AU. $r_\mathrm{out}$ and $M_{disk}$ can be constrained by our observations assuming a dust opacity (see Section 4.2.3) and gas-to-dust mass ratio of 100.
\subsubsection{Modeling the Envelope}

We adopted the rotating and collapsing spheroid structure by \citet{Ulrich1976} to model the envelope. The density of this envelope structure is given by,
   \begin{equation}
      \mathrm{\rho_{\rm env}}(r,\mathrm{\theta}) = \mathrm{\rho_{0}}\left( \frac{R_{\rm rot}}{r} \right)^{3/2} \left(1 + \frac{\rm cos \mathrm{\theta}}{\rm cos \theta_{0}} \right)^{-1/2} \left( \frac{\rm cos \theta}{2 \rm cos \theta_{0}} + \frac{R_{\rm rot}}{r}\rm cos^{2}\theta_{0} \right)^{-1} \,,
   \end{equation}
where $\rho_\mathrm{0}$ is the density in the equatorial plane at the centrifugal radius $R_\mathrm{rot}$ of the envelope, and $\theta_\mathrm{0}$ is the solution of the parabolic motion of an infalling particle given by:
\begin{equation}
\frac{r (\mathrm{cos}\theta_{0} - \mathrm{cos}\theta)}{(R_\mathrm{rot} \mathrm{cos}\theta_{0} \mathrm{sin}^{2}\theta_{0})} = 1 \,
\end{equation}
The outer radius of the envelope is fixed at 8\,800 AU, which is equivalent to the 30$^{\prime\prime}$ aperture of Bolocam. In this case we will use the envelope mass derived by Bolocam to compare with the models. 
We computed $\rho_{0}$by imposing a total envelope mass M$_\mathrm{env}$, and $R_{rot}$, which can have a
significant influence on the amplitude as a function of baseline, it was left free to vary. Outflow cavities are not included in this model. \\
RADMC-3D is used to compute the temperature of the envelope, with the implementation of Eq. (5) to describe the density structure. The protostar and disk system presented in the previous subsections are used as heating source of the envelope, whose emission is calculated using the two-layer model by \citet{Dullemond2001}, and then the output spectrum is used in the 2D radiative transfer calculation for the envelope structure. 
\subsubsection{Dust opacity}
We adopt the dust opacity model used in \citet{Ricci2010}. A dust population characterized by a distribution of grains with different sizes was implemented. 
We used a truncated power law distribution n($a$) $\propto$ $a$$^{-q}$, between a minimum and a maximum grain size, $a_\mathrm{min}$ and $a_\mathrm{max}$ respectively. We fixed the chemical composition to a silicate, carbonaceous material and water ice in a 1:2:3 volume fractional ratio. Additionally, we set $a_\mathrm{min}$=0.01 $\mu$m and we use \textit{q}=3.0. We varied $a_\mathrm{max}^\mathrm{disk}$ and $a_\mathrm{max}^\mathrm{env}$ according to the range presented in Table 5.

\begin{table*}
\caption{Two--step Model best-fit parameters}  
\label{bestfit2step}
\centering          
\begin{tabular}{c c c c}     
\hline\hline      
Parameter & Description & Best--fit M1 & Best--fit M2 \\
\hline
Disk model parameters \\
$\mathit{R_\mathrm{out}}$ (AU) & Disk outer radius & 32 & 34 \\
$\mathit{M_\mathrm{disk}}$ ($M_{\odot}$) & Disk mass & 0.4 & 0.2 \\
$\mathit{\Sigma_\mathrm{disk}}$ (gr $\mathrm{cm^{-2}}$) & Disk surface density & 554.13 & 245.38\\
$\mathit{a_\mathrm{max}^\mathrm{disk}}$ ($\mu$m) & Disk maximum grain size & 10\,000 & 10\,000  \\
RADMC-3D / Envelope parameters \\
$\mathit{R_\mathrm{rot}}$ (AU) & Centrifugal radius & 600 & 600\\
$\mathit{\rho_\mathrm{0}}$ (gr $\mathrm{cm^{-3}}$) &  Density in the equatorial plane at $R_\mathrm{rot}$ & <8.5$\times10^{-20}$ & <6.0$\times10^{-20}$ \\
$\mathit{a_\mathrm{max}^\mathrm{env}}$ ($\mu$m) & Envelope maximum grain size & <100 & <100 \\
\hline
\end{tabular}
\end{table*}

\subsubsection{Model fitting}

\begin{figure*}[h!]
\centering
\includegraphics[width=8cm]{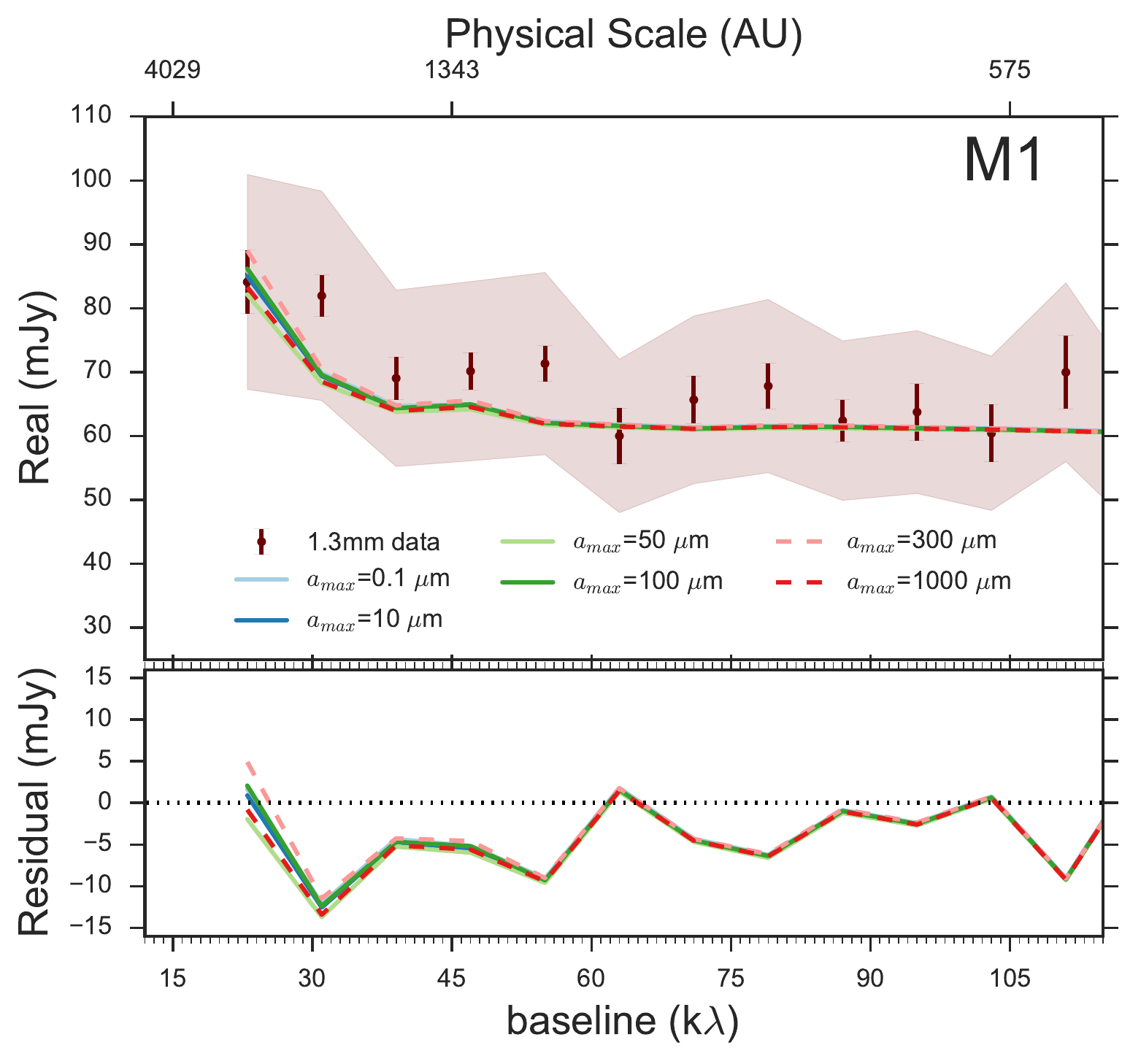}
\includegraphics[width=8cm]{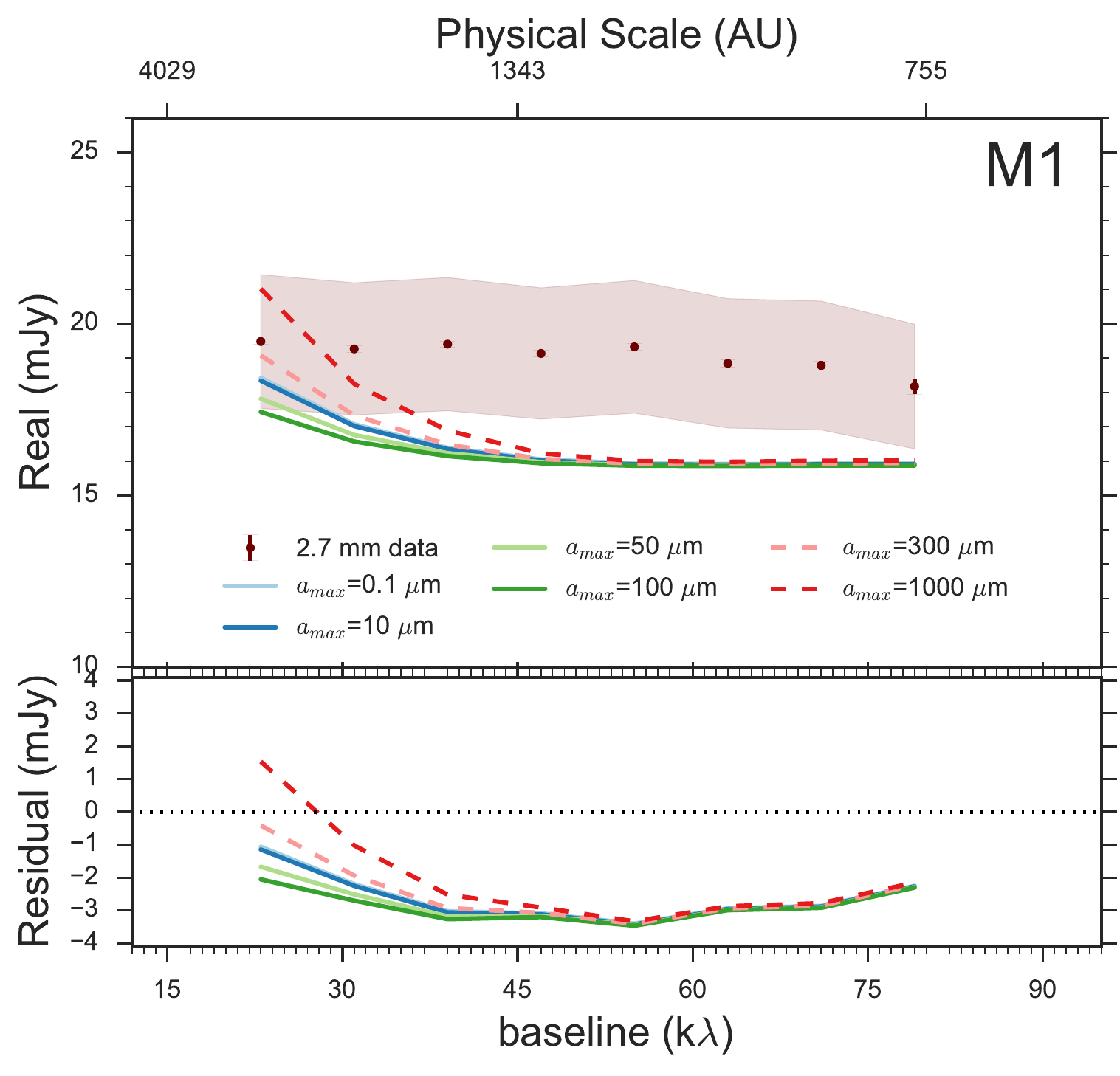}
\includegraphics[width=8cm]{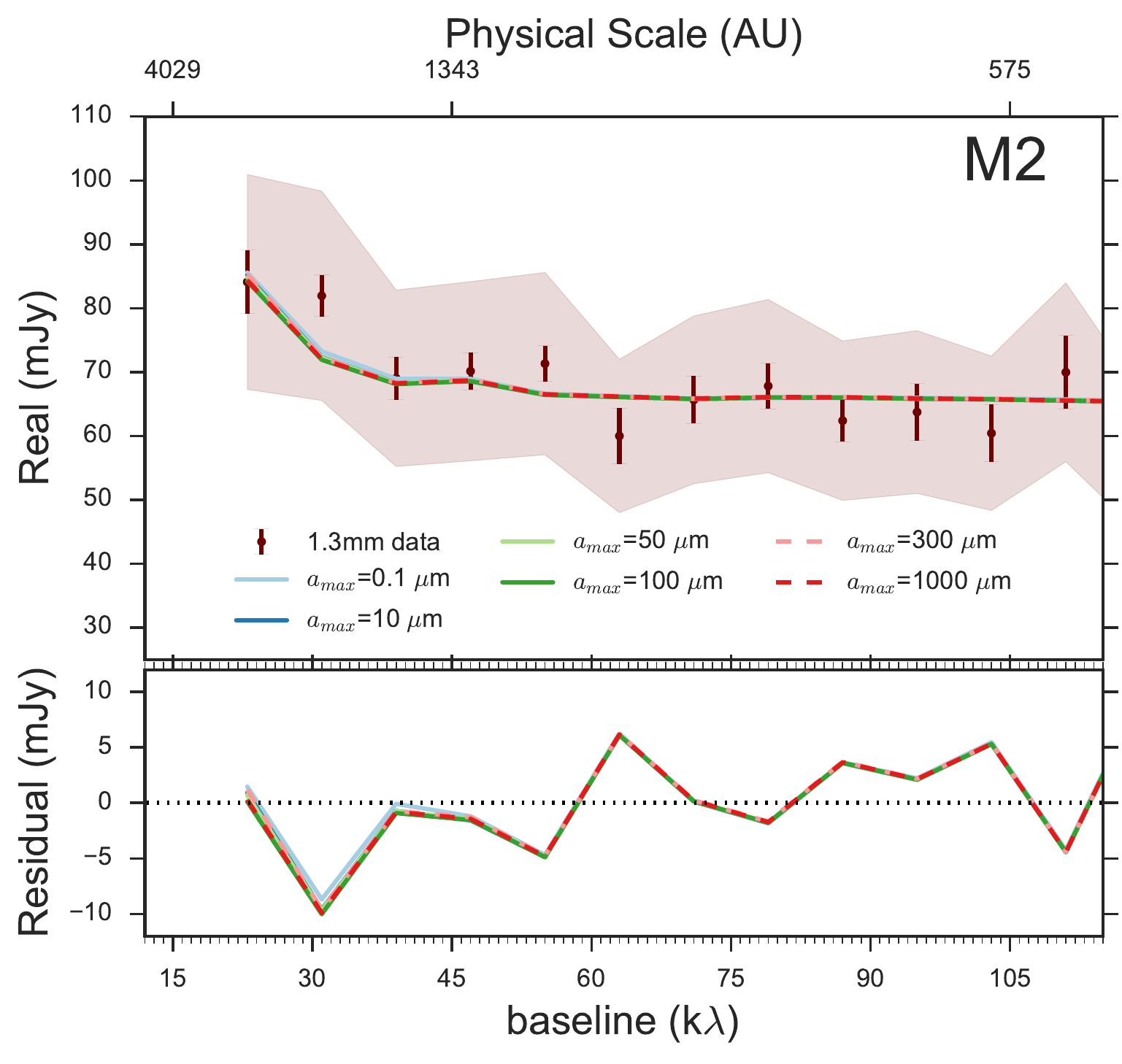}
\includegraphics[width=8cm]{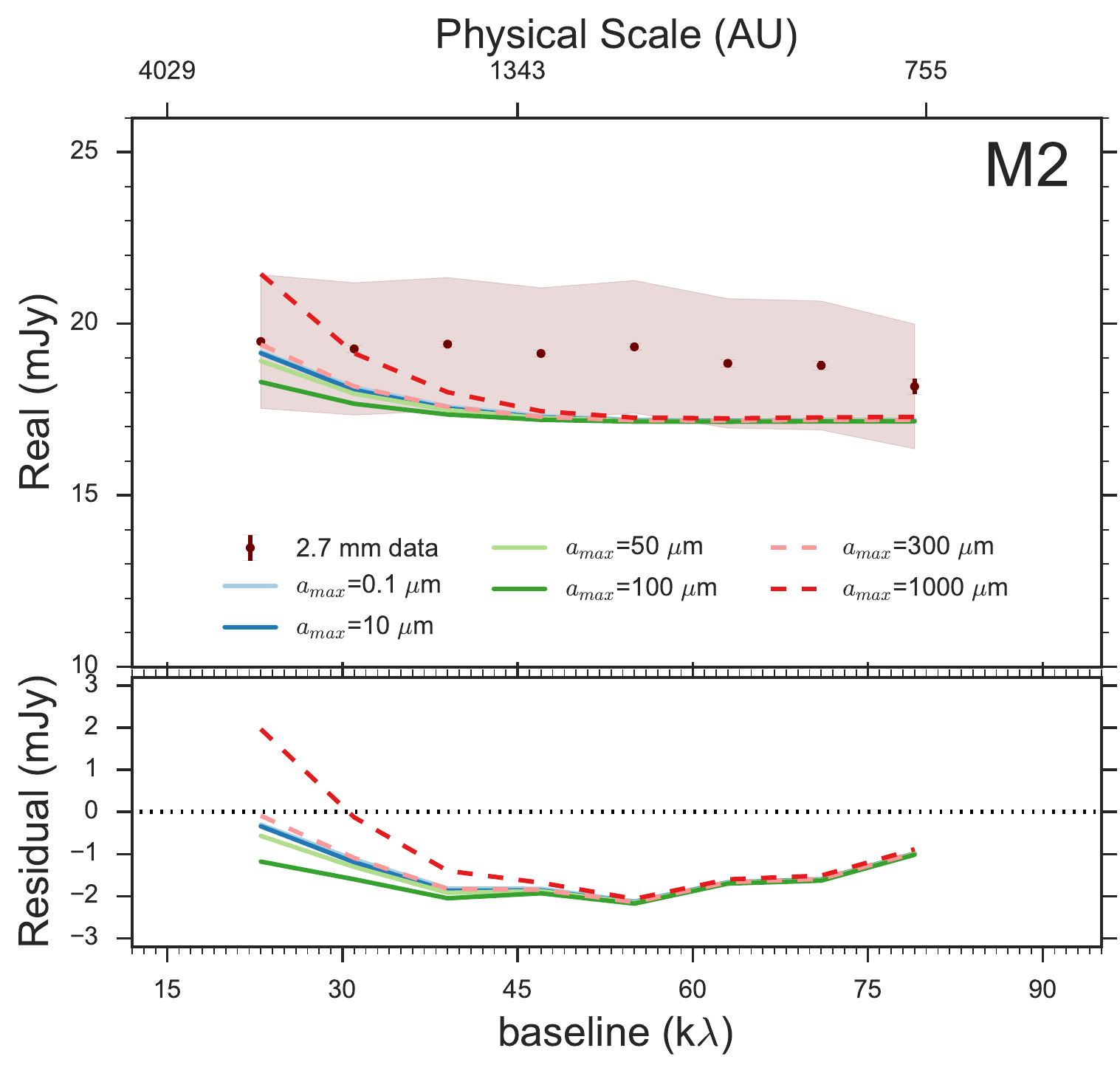}
\caption{Real part of the visibilities as a function of baseline. Left panels are 1.3 mm data while right panels are 2.7 mm data. The two upper panels are models with the disk model M1 while the bottom panels are presented the models using disk model M2 (see Table \ref{bestfit2step}). In solid lines we present models with grain sizes of $a_\mathrm{max}\leq100$ $\mu$m. In dashed lines are models with grain sizes of $a_\mathrm{max}=$300,1000 $\mu$m. The best fits are the models with a distribution of grain sizes with $a_\mathrm{max}\leq100$ $\mu$m. The red shaded region is the uncertainty on the data due to flux calibration. The bottom of each panel shows the residuals between the data and the model with different $a_\mathrm{max}$.}
\label{2stepfigures}
\end{figure*}

To compare the model with the interferometric observations, we have to create images at the exact wavelengths of our observations. Then, those model images have to be transformed to model visibilities. For that we used the computational library GALARIO \citep{Tazzari2018}. The model image is convolved with the primary beam patterns of the antennas and then Fourier transformed into visibilities.
\\
The first step in this modeling is to fit the disk emission. We created a grid of parameters varying M$_\mathrm{disk}$, $R_\mathrm{out}$ and $a_\mathrm{max}$ to reproduce together $F_\mathrm{disk}^\mathrm{1.3mm}$ and $F_\mathrm{disk}^\mathrm{2.7mm}$.
Once we found the three parameters that match F$_\mathrm{disk}^\mathrm{1.3mm}$=63.97 mJy and F$_\mathrm{d}^{2.7\mathrm{mm}}$=18.8 mJy, we implement these output fluxes (output spectrum) 
as the heating central source of the envelope. 
\\
Then, using RADMC-3D \citep{Dullemond2012}, we vary $M_\mathrm{env}$ and $a_\mathrm{max}^\mathrm{env}$ in order to reproduce the interferometric fluxes at 1.3~mm and 2.7~mm. 
Table \ref{grid2step} gives a complete list of models parameters and indicates whether they are fixed or varied.
In Fig. \ref{2stepfigures} we present the best fit for the observed visibilities at both wavelengths. The set of parameters that provided the best match with the observations are presented in Table \ref{bestfit2step}. The two best fit are discussed in the next section.
\subsubsection{Results Two step model}
\begin{figure*}
\centering
\includegraphics[width=16cm]{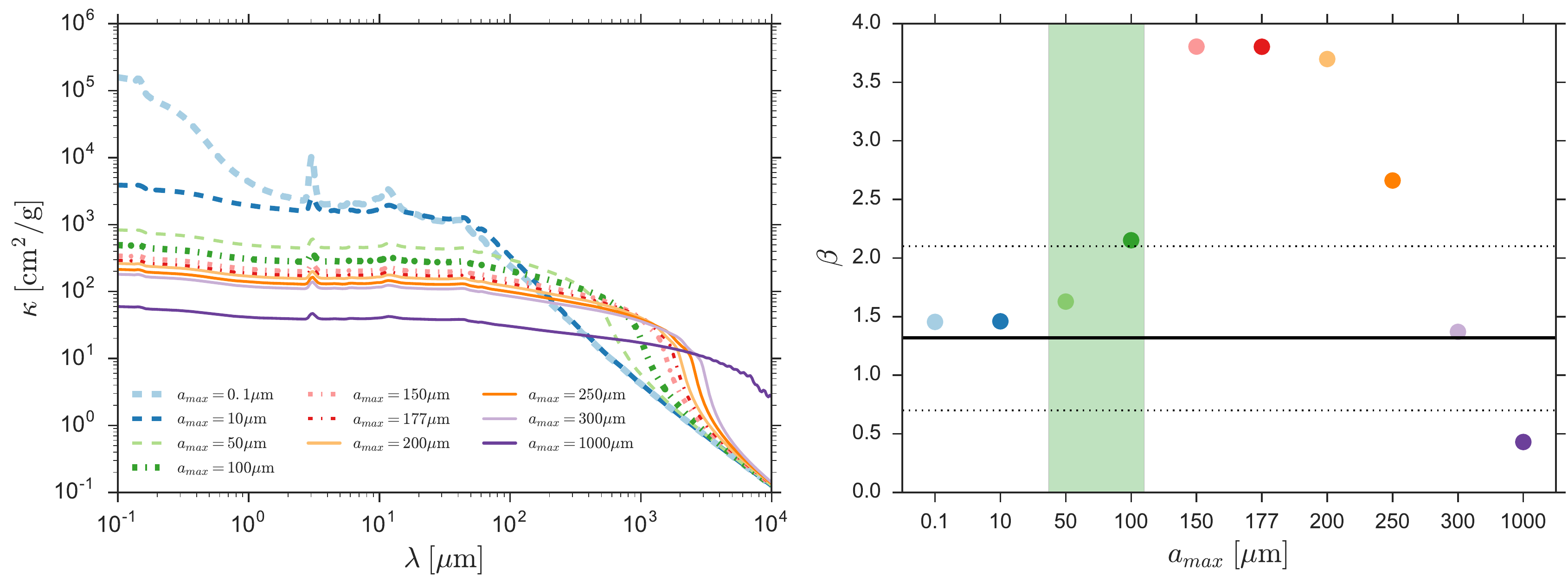}
\caption{Left panel shows the dust absorption opacity as a function of wavelength for grain size distributions characterized by ($a$) $\propto$ $a^{-3.0}$ and increasing maximum grain size ($a_{max}$). Right panel shows the dust opacity spectral index ($\beta$) calculated between 1.3~mm and 2.7~mm wavelengths as a function of the maximum grain size. Black solid line is the $\beta_\mathrm{env}$ value from the parametric model and the black dashed lines are the uncertainties. Green region shows an upper limit for $a_\mathrm{max}$ in the envelope of Per-emb-50.}
\label{opacities}
\end{figure*}

The parameters that provide a good fit respect to the disk emission at both wavelengths are reported in Table \ref{bestfit2step}. The model M1 with a 32 AU disk radius and $M_\mathrm{disk}=0.4\rm M_{\odot}$ is consistent with the rescaled values reported by \citet{SeguraCox2016}. 
While all the disk models match the long baselines 1.3~mm data, the disk emission at 2.7~mm is 15\% lower than the data.
On the other hand, the disk model M2, with a 34~AU disk radius and $M_\mathrm{disk}=0.2 \rm M_{\odot}$ matches very well the observations at both wavelengths, but compared with values of Table \ref{literature}, the disk radius is slightly larger. \\
The differences in the disk models may be due to the assumed values of $\kappa_\mathrm{\nu}$= 0.00146 $\mathrm{cm}^{2}$ $\mathrm{g}^{-1}$ and disk temperatures of 20~K and 40~K in \citet{SeguraCox2016}. Therefore, higher resolution millimeter observations that would resolve the disk are needed to put much stronger constraints on Per-emb-50.
\\
\\
For the envelope, we explore the effects of changing: $R_{rot}$, $a_\mathrm{max}$ and $\rho_{0}$. The envelope inner radius is fixed at the outer radius of the disk model. We tested different $R_{rot}$ between 100-1000\,AU to accommodate the total enclosed envelope mass. As mentioned in \citealt{Crapsi2008}, decreasing the centrifugal radius results in more peaked and spherical envelopes. Using a small centrifugal radius has a significant influence on the amplitude at short baseline length. For example varying the centrifugal radius by a factor of 2 changes the first amplitude point of the model by 20\%. We found that a $R_\mathrm{rot}$ of 600 AU is consistent with the slope at short baselines in both wavelengths. \\
We can constrain the level of grain sizes in the envelope within the framework of the collapsing rotating envelope model. For example, in Fig. \ref{2stepfigures}, if we consider a dust grain size distribution in the envelope with a maximum size of 1\,mm, we can reproduce the 1.3\,mm observations, but we underestimate the total envelope mass by a factor of 6.  
In the case of models with 0.1\,$\mu$m< $a_\mathrm{max}$ $<100$\,$\mu$m, the flux at 1.3\,mm and 2.7\,mm matches the observations very well, but the derived envelope masses differ from those derived from observations. 
The best match with the $2.2 \rm M_{\odot}$ envelope mass derived by \citet{Enoch09} are those derived from models with dust grain sizes of $a_\mathrm{max}\leq50\mu\mathrm{m}$ (see Table \ref{bestfit2step}). The models with $a_\mathrm{max}=100~\mu\mathrm{m}$ recover almost 60\% of the envelope mass. Table \ref{envelopemass-table} presents the derived masses for the envelope using different $a_\mathrm{max}$ in M1 and M2. \\
A distribution of grains with $a_\mathrm{max}$$\leq$$50$ $\mu \mathrm{m}$ provides a good match with the observations since the flat emission at 2.7\,mm matches the observations well and is consistent with the systematic errors due the flux calibration. \\
Based on this model, the maximum grain sizes in the envelope are unlikely to be larger than a hundred microns. This would imply that the envelope may have gone through a process of grain growth, but there is no evidence that a substantial fraction of grains are large millimeter-sized dust aggregates. \\
As we mention before, the observed flux and spectral index of Per-emb-50 are consistent with a small optically thick disk, in which case, we cannot constrain the spectral index $\alpha$. \\
For the envelope we can use our dust model to infer the value of $\beta$, which is $\beta_\mathrm{env}$=1.46 and $\beta_\mathrm{env}$=1.63 for $a_\mathrm{max}$=10 and $a_\mathrm{max}$=50 $\mu$m, respectively. In Fig. \ref{opacities} we compare the different $\beta$ values for each $a_{max}$ with the value obtained from the parametric model. The $\beta$ values for 0.1$\mu$m<$a_\mathrm{max}$<100$\mu$m are consistent, within the uncertainties, with the $\beta$ calculated with the parametric model. In the case of grains larger than 100 $\mu$m, the total envelope mass is underestimated.
\begin{table}[h]
\caption{Derived envelope masses}             
\label{envelopemass-table}      
\centering                          
\begin{tabular}{c c c}        
\hline\hline           
Model & $a_\mathrm{max}$ ($\mu\mathrm{m}$) & $M_\mathrm{env}$ ($M_{\odot}$)  \\    
\hline\hline                        
M1 &  0.1,10 & 1.73     \\
 &  50 & 1.53     \\
 &  100 & 1.15     \\
  &  300 & 0.23     \\
 &  1000 & 0.34     \\
\hline
M2 &  0.1 & 1.22     \\
 &  10,50 & 1.38  \\
 &  100 & 0.85     \\
  &   300 & 0.16     \\
  &   1000 & 0.29     \\
\hline                                   
\end{tabular}
\tablefoot{Envelope mass calculated within 8\,800 AU radius}
\end{table}
\subsection{Full radiative transfer model}
In this model we used a system that consists of disk, protostar, envelope and outflow cavity. We used the radiative transfer tool RADMC-3D from \cite{Dullemond2012} to compute the emission from all the contributions. The details of each contribution will be discussed in the following sections. 
\subsubsection{Disk model}
We adopt a disk model heated by its protostellar radiation. The surface density profile $\Sigma(R)$ was modeled as a truncated power law as in Eq. 4, with a power exponent of the surface density distribution \textit{p}=1; $\Sigma_{0}$ is scaled to accommodate the total mass of the disk $M_\mathrm{disk}$.
The 2D volume density with an exponential vertical profile is defined by:
\begin{equation}
\rho(r,z) = \frac{\Sigma(r)}{H_{p}\sqrt{2\pi}} \exp{\left(-\frac{z^2}{2H_{p}^2}\right)} \ ,
\end{equation}
where \rm $H_{p}$ is the pressure scale height and is defined as $H_{p}/r$=0.1(r/$r_{h_{p}}$)$^{\phi}$, $r_{h_{p}}$ is the reference radius set at 25 AU, and $\phi$ is the flaring index of the disk, which in this case is set to 1.14, as an average value according to previous studies on young sources \citep{Pineda2011,Tobin2013}. 
We used the disk inclination angle, disk radius, and disk mass presented in Table \ref{literature}.
\begin{table*}[t]  
\centering
\caption{Full radiative transfer model grid parameters}  
\label{gridfullradmc}
\begin{tabular}{c c c c}     
\hline\hline      
Parameter & Description & Values & Parameter Use \\
\hline
Stellar model parameters  \\
$M_{\star}$ ($M_{\odot}$)& Stellar mass & 2.9 & fixed\\
$R_{\star}$ ($R_{\odot}$) &  Stellar radius & 5.0 & fixed\\
$T_{\star}$ (K) & Effective temperature & 5011 & fixed \\
\hline
Disk parameters \\
$\Sigma_{bkg}$ (gr $\mathrm{cm^{-3}}$) & Background density & $1.0\times10^{-30}$ & fixed\\
$M_\mathrm{dust}/M_\mathrm{gas}$&Dust-to-gas mass ratio & 0.01 & fixed\\
$R_{H_{p}}$ (AU)& Reference radius at which $H_{p}/R$ is taken & 25 & fixed\\
$m_\mathrm{disk}$ ($M_{\odot}$) &Mass of the disk & 0.18--0.36 & varied\\
$\phi$ & Flaring index & 1.14 & fixed \\
$p$ & Power exponent of the surface density distribution& 1.0 & fixed\\
$r_\mathrm{out}$ (AU) & Disk outer radius & 25,27,30,32 & varied\\
$r_\mathrm{in}$ (AU) & Disk inner radius & 1.0 & fixed \\
$a_\mathrm{max}^\mathrm{disk}$($\mu$m) &Disk maximum grain size& 10000 & fixed\\
\hline
Envelope parameters \\
$R_\mathrm{out}$ (AU) & Envelope outer radius & 8,800 & fixed\\
$\alpha$& Power exponent of the radial density distribution & -1.1,-1.5,-1.8 & varied\\
$n_{0}$ (gr $\mathrm{cm^{-3}}$) &  Central density  & $1.0,1.5,2.0\times10^{-16}$ & varied\\
$r_\mathrm{0}$ (AU)& Within this radius the density profile is flat & 25,27,30,32 & varied\\
$\theta$ ($^{\circ}$)&Opening angle of the outflow&30 & fixed\\
$a_\mathrm{max}^\mathrm{env}$($\mu$m) &Envelope maximum grain size& 0.1--1000 & varied\\
\hline
\end{tabular}
\tablefoot{Each model is calculated with 1$\times10^{6}$ photons for the thermal Monte Carlo.}
\end{table*}
\subsubsection{Envelope model}
For the envelope model we adopted a density profile by \citet{Tafalla2002}, which combines a power-law behavior for large radius and a central flattening profile at small radius, i.e.,
\begin{equation}
n(r) = \frac{n_{0}}{1+(r/r_{0})^\alpha} \,,
\end{equation}
where $n_{0}$ is the central density, $r_{0}$ is the radius of the flat region or truncation radius, and $\alpha$ is the asymptotic power index. The outer radius of the envelope is fixed at 8\,800 AU to match the beam of \citet{Enoch09} observations, in which the rescaled envelope mass is 2.2 \rm M$_{\odot}$. Additionally, since we have evidence of an outflow in this source \citep{Ian2017}, we included an outflow cavity with an opening angle of 30$^{\circ}$ (M. Dunham, priv. comm.) and a lower density of $1.0\times10^{-30}$ $\mathrm{gr \, cm^{-3}}$ for the region inside the cavity and the background.

\subsubsection{Backwarming effect}

The effects of the envelope thermal emission on disk (i.e., backwarming) have been studied in different environment, as in the case of the heavily embedded source L1551 IRS 5 \citep{Butner1994}. 
\\In the case of an envelope around a disk, the millimeter emission of the disk increases. This is because the envelope acts as a thermal cavity, not letting the temperature within the cavity to fall below the temperature of the envelope wall. Therefore, a substantial backwarming effect on the disk can be present depending on the optical depth and geometry of the cavity.
\\
In the previous envelope modeling following \citet{Miotello14}, this effect has been ignored due the geometry of the envelope. Different profiles might heat the disk to a different degree.
To explore the effects of backwarming we have computed new models which attempt to take it into account. The net effect of the envelope on the disk temperature is discussed in the Appendix D. 

\subsubsection{Dust opacity}


We used two kind of dust opacities in order to test the model. Firstly, we used the opacity computed in \citet{Ossenkopf1994} based on a coagulated grain size distribution. In this model, a truncated power law is adopted for the initial dust distribution, n($a$) $\propto$ $a^{-q}$, where the minimum size of the grain is $a_\mathrm{min}$=5 nm, the maximum size is $a_\mathrm{max}$=250 nm and the power index \textit{q} is set to 3.5. The dust distribution is calculated after 10$^{5}$ years of coagulation with a gas density of $n_{\rm H}$=10$^{5}$ cm$^{-3}$ expected in a prestellar core. 
Secondly, we used the previous dust opacities presented in Section 4.2.3. \\ Since the second dust opacity approach covers maximum grain sizes from small grains of 0.1$\mu$m, to big grains of 1\,cm, we decided to present here the results with those opacities to compare consistently with the previous modeling.
\subsubsection{Model fitting}
The free parameters for the disk are the outer radius, $r_\mathrm{out}$ and the disk surface density $\Sigma_{0}$. The free parameters for the envelope are its mass $M_{env}$, its power law density profile $\alpha$, its flattening envelope radius $r_\mathrm{0}$ and its dust opacity, characterized by $a_\mathrm{max}^\mathrm{env}$. The truncation radius of the envelope is set at the outer radius of the disk parameter. \\
Since the disk parameters estimated by \citet{SeguraCox2016} are not solid constraints, we test our model using their mass and outer radius values as an upper and lower limit on $\Sigma_\mathrm{0}$. The grid of parameters that we test and set are presented in Table \ref{gridfullradmc}.
Once the dust temperature of the system is calculated from the input parameters of Table \ref{gridfullradmc}, we compute the synthetic images, for 1.3mm and 2.7mm, following the same procedure reported in Section 4.2.4.
We simultaneously fit the 1.3mm and 2.7mm visibilities by calculating $\chi^{2}$ values for each model using the equation
\begin{equation}
\chi^{2} =\sum_{i=1}^N \frac{(F_{\nu,\mathrm{observed},i} - F_{\nu,\mathrm{model},i})^{2}}{\sigma^{2}_{i}},
\end{equation}
for the entire set of visibility points between 20 and 110 k$\lambda$. The uncertainty in the data, $\sigma_{i}$, includes the statistical uncertainty and the absolute flux uncertainty of 10\% for 2.7~mm data and 20\% for 1.3~mm data, both added in quadrature.
Since our observational constraints are dominated by the errors of the data sets, it is possible that the disk and/or envelope structure would be wrong at some level, therefore, our $\chi^{2}$ value is simply an indicator of an acceptable model, not a best fit. After performing a visual inspection of the models, we report the best match with the observations in the next paragraph and in Table \ref{bestmodelradmc}. \\
A sample of models with different $a_{max}$ and derived envelope masses are presented in the Appendix \ref{144models}.

\begin{figure*}[h]
\centering
\label{fullfigures}
\includegraphics[width=6.95cm]{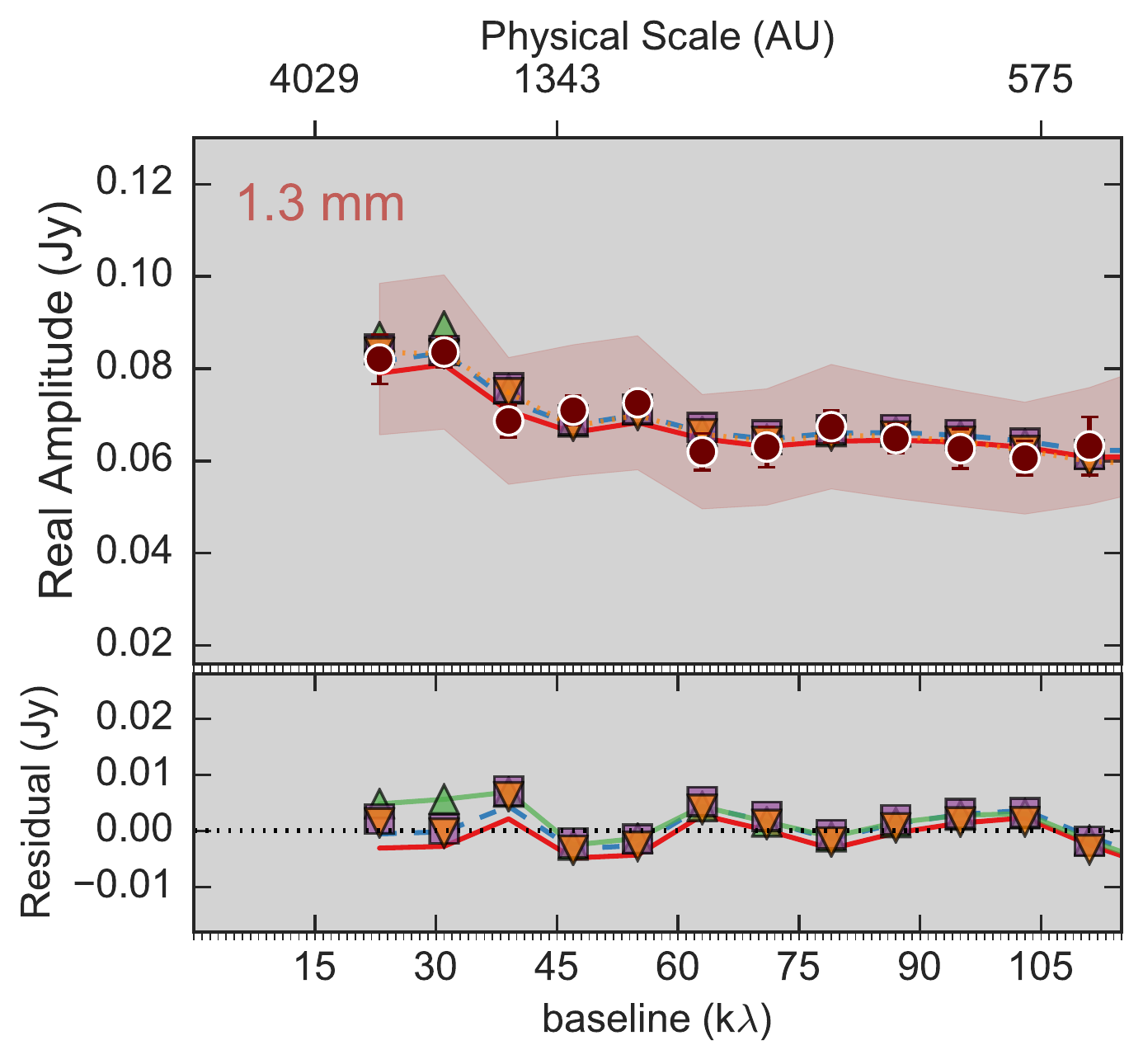}
\includegraphics[width=11.15cm]{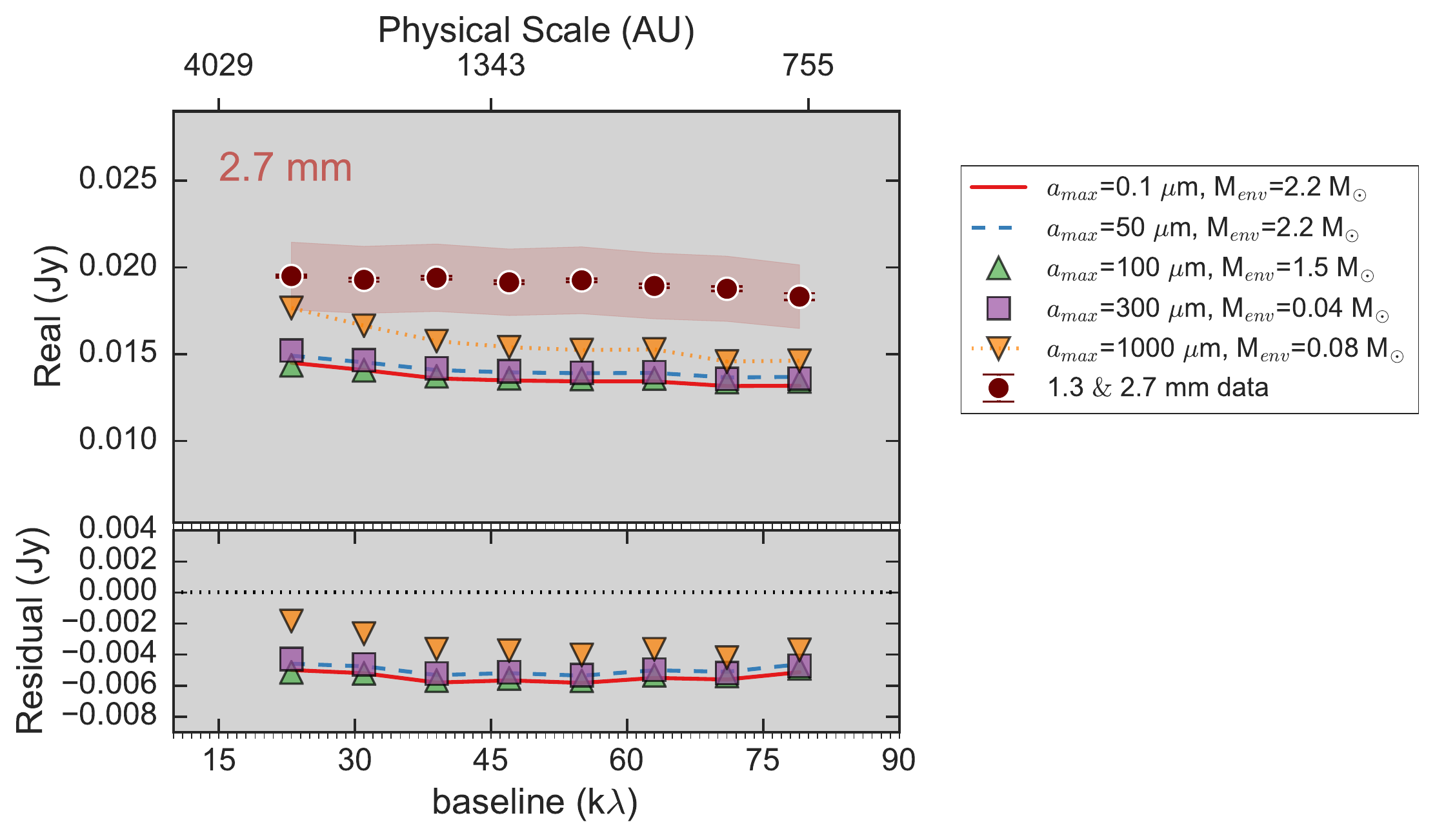}
\caption{Real part of the visibilities as a function of the deprojected baseline. Left panel shows 1.3 mm data while right panel shows 2.7 mm data. Red shaded regions are the uncertainties due the flux calibration. We show a variety of models with a maximum grain size in the envelope of $a_{\rm max}=0.1,50,100,300,1000$ $\mu$m. At the bottom of each panel are the residuals between the data and the best model.}
\end{figure*}

\subsubsection{Results full radiative transfer model}

\begin{table*}  
\centering
\caption{Full radiative transfer best fit models}  
\label{bestmodelradmc}
\begin{tabular}{c c c c}     
\hline\hline      
Parameter & Description & Best-Model 1 & Best-Model 2 \\
\hline
Disk parameters \\
m$_\mathrm{disk}$ ($M_{\odot}$) &Mass of the disk & 0.20 & 0.24\\
$\Sigma_\mathrm{disk}$ (gr $\mathrm{cm^{-2}}$) & Disk surface density & 362.2 & 364.2\\

$r_\mathrm{out}$ (AU) & Disk outer radius & 25 & 27\\
$a_\mathrm{max}^\mathrm{disk}$($\mu$m) &Disk maximum grain size& 10\,000 & 10\,000\\
\hline
Envelope parameters \\
$\alpha$& Power exponent of the radial density distribution & -1.1 & -1.1\\
$\rho_{0}$ (gr $\mathrm{cm^{-3}}$) &  Central density  & $2.0\times10^{-16}$  & $1.5\times10^{-16}$\\
$M_\mathrm{env}$ ($M_{\odot}$) &  Envelope mass  & 2.24 & 1.54 \\
$r_\mathrm{trun}$ (AU) & Truncation radius & 25 & 27 \\
$r_{0}$ (AU)& Within this radius the density profile is flat & 25 & 27\\
$a_\mathrm{max}^\mathrm{env}$($\mu$m) &Envelope maximum grain size& 50 & 100\\
\hline
\end{tabular}
\end{table*}

From our interferometric observations, we are limited to study the inner regions of the envelope, from 4\,000 AU to 600 AU, therefore, we examine a power-law density profile follow \citet{Tafalla2002}. An unresolved component is included to represent a compact disk structure.
In the envelope we used a constant dust grain population, in which we vary the maximum grain size from 0.1 $\mu$m to 1\,000 $\mu$m.
To study the impact of the maximum grain size in the envelope, $a_\mathrm{max}^\mathrm{env}$, the central density, $n_{0}$, and the density power law index, $\alpha$, we used the range of parameters reported in Table \ref{gridfullradmc}.

Table 9 shows the model parameters that provides the best fits to the observations.
For the disk properties, we compare our results with the values reported in Table \ref{literature}. Our disk mass and radius are consistent with the rescaled values reported by \citet{SeguraCox2016}. 
Both disk models in Per-emb-50 are consistent with a small optically thick disk, but do not allow us to probe if there is grain growth throughout the disk since we are missing very long baselines to resolve the disk.

Similar to the results of the two-step model, the full radiative transfer models suggest a distribution of dust grains in the envelope with maximum size $a_\mathrm{max}^\mathrm{env}=50,100$ $\mu$m and a resulting envelope mass within 8\,800 AU radius of $M_\mathrm{env}$$\sim$2.24, 1.54 M$_{\odot}$, respectively. In Fig. 8, we present a variety of models with different $a_{\rm max}$ in the envelope that match the visibility data. While all the models match the 1.3\,mm data within the flux uncertainty (red region), the 2.7\,mm data allow us to determine a good model because the shape of the short baseline emission. \\
Models with $a_\mathrm{max}^\mathrm{env}$ < 100 $\mu$m follow the flat emission of the 2.7\,mm data, while models with $a_\mathrm{max}^\mathrm{env}$ > 300 $\mu$m overestimate the short baseline emission at 2.7mm and underestimate the envelope mass of Table \ref{literature}. These results are consistent with our previous modeling and in agreement with the spectral index $\beta$ that we calculated in Fig. \ref{opacities}. We also reported the 1.1\,mm single dish flux (see Table \ref{1.1fluxes}) for each model. 

\begin{table}[h!]
\caption{Derived 1.1\, ${\rm mm}$ fluxes \& envelope mass}             
\label{1.1fluxes}      
\centering                          
\begin{tabular}{c c c}        
\hline\hline                 
$a_\mathrm{max}$ ($\mu$m) & $M_\mathrm{env}$ ($M_\mathrm{\odot}$) & $F_\mathrm{1.1\,mm}$ (Jy) \\ 
\hline                        
   0.1 & 2.24 & 1.87 \\
   50 & 2.24 & 1.76    \\
   100 & 1.54 & 3.18    \\
   300 & 0.04 & 0.37    \\
   1000 & 0.08 & 0.35    \\
\hline                                   
\end{tabular}
\tablefoot{Each mass model is calculated within a 8\,800 AU envelope radius. The 1.1\,mm fluxes are calculated using an aperture of 30", simulating the diameter aperture of Bolocam.}
\end{table}
As discussed by many authors \citep{Draine2006, Banzatti2011,Testi2014}, it is quite difficult to explain values of $\beta$ less than 1 without invoking the presence of millimeter size grains, regardless of the chemical composition, porosity, or grain geometry. 
In the case of Per-emb-50, the high value of $\beta$ is compatible with grains no larger than 100 $\mu$m, and with values found in Class 0 sources by \citet{Jennifer2017}.
The fact that we find grains that have not reached mm sizes in the envelope of Per-emb-50 will be discussed in the following section.
\section{Discussion}
\subsection{Grain sizes in Class I protostellar envelopes}
The presence of millimeter-size grains in envelopes of young protostars, Class 0/I, have been studied and modeled by many authors (e.g \citealt{Kwon09,Chiang2012, Tobin2013, Miotello14}), but current models cannot easily explain growth at that level \citep{Ormel:2009dq}. This is because the models require high number densities, $n_{\mathrm{H}}$ > $10^{6}$ cm$^{-3}$, to form such large grains in timescales of 1 Myr. \\
\citet{Miotello14} found that dust grains start to aggregate up to mm sizes already in the envelope of two Class I protostars, producing a change in the spectral index with values of $\beta_{\mathrm env}$=0.6$\pm$0.3 for Elias 29 and $\beta_{\mathrm env}$=0.8$\pm$0.7 for WL 12. Those values are smaller than the spectral index for the envelope of Per-emb-50, $\beta_{\mathrm env}$=1.4$\pm$0.3, by a factor of 2. The differences between these studies may be associated to the properties of the star forming region. In our case Per-emb-50 is in NGC1333 region in Perseus, which is a very crowded region with young stellar objects, while the sources of \citet{Miotello14} are isolated and embedded in L1688 in Ophiuchus. With this study on Per-emb-50, we suggest the possibility that: (a) millimeter grains in envelopes of young protostars may not be a common result, or (b) the dust grain growth is not a homogeneous process. Finally, the environment within which a protostar forms could also play a role in the amount of dust coagulation, which significantly affects the future formation and structure of the protoplanetary disk, as shown by \citet{Zhao2016, Zhao2018}. \\ 

The possibility that grains can grow up to mm-size in the envelope of Class 0/I protostars was studied by \citet{Wong2016}. They proposed another mechanism to explain the existence of mm-size grains in the envelopes of young protostellar sources that consists of transport of mm-sized grains from dense regions close to the protostar to the envelope via the outflow. This scenario is quite plausible before the central mass of the protostar reaches a mass of 0.1 $M_\mathrm{\odot}$ with a mass-loss rate of 10$^{-6}$ $M_\mathrm{\odot}$ yr$^{-1}$. This could be the case of Per-emb-50, but high resolution data is needed to model the inner regions of this source.
\\ The results of our analysis show that dust grains may have grown as large as $\sim$100 $\mu$m in size in the envelope of Per-emb-50 at scales of 4000-2000 AU. This implies that there is a degree of grain growth with respect to the ISM sizes, but not significant enough to lower the value of $\alpha$. This is also in agreement with the work of \citet{Chacon2017}, where they predict grain sizes of a few hundred $\mu$m in the central 300 AU of the prestellar core L1544.

Taking this into account, it is crucial to perform surveys for Class I protostars embedded in different environments and at different physical scales to determine the variation of $\alpha$ spectral index and the corresponding amount of grain growth. 

\subsection{The effects of Backwarming}
We find that backwarming is important for modeling Per-emb-50. From the previous analysis, using the two--step modeling, it was straightforward to fit the nearly constant emission at long baselines with an unresolved disk. For the full radiative transfer modeling, this was not the case. Considering the full radiative transfer model, the use of a \citet{Tafalla2002} density profile combined with a power-law behavior for large radius and a central flattening profile at small radius shows that backwarming is important since the disk emission is completely affected by the addition of the envelope. This change in emission is discussed in \citet{Butner1994} and in the Appendix \ref{appfig}. To study the effects of different envelopes geometries on disk emission is beyond the scope of this work. 
\\
However, backwarming can have other consequences. The change of temperature between a backwarmed disk ($\sim$100 K) and a nonbackwarmed disk with $\sim$20 K, would affect significantly the gas phase chemistry and the dust mantle chemistry in young disks and envelopes \citep{Butner1994}. Finally, the backwarming in Class I protostars could have an important effect in the thermal history of the outer disks of planetary systems. Detailed studies using proper physical structures and radiative transfer models are necessary to address the backwarming effect present in most young embedded sources.

\section{Conclusions}

We present new 1.3~mm data from SMA and 2.7~mm data from NOEMA of the brightest Class I protostar Per-emb-50 in the NGC 1333 cluster in the Perseus star forming region. 
In the u--v plane it is possible to distinguish the presence of a large scale envelope at short baselines and an unresolved and optically thick disk at longer $u$-$v$ distances. From the data analysis and the different modeling approaches on this source we can conclude:

\begin{itemize}
\item For the envelope $\mathrm{uv}$ analysis we find a spectral index similar to the typical ISM values, $\alpha_\mathrm{mm}$=3.3$\pm$0.3.
\item The current observations on Per-emb-50 and the radiative transfer modeling reveal a Class I envelope consistent with maximum sized grains of < 100 $\mu$m. This suggests that grain growth has proceeded within the envelope, but not to a level to produce changes in $\alpha$ as the presence of millimeter size grains does.
\item The presence of grains with a size range of <100 $\mu$m in envelopes of Class I protostars may have an impact in our understanding of protostellar evolution. Following the prediction from \citet{Chacon2017}, who find that dust grains are expected to grow to sizes of a few hundred $\mu$m in the central 300 AU of a pre-stellar core, we could suggest that the larger grains found in the envelope of Per-emb-50 may be inherited from the prestellar phase.
\item These results show for the first time no evidence of grain growth to millimeter sizes in the inner regions of the envelope of a Class I protostar, providing an interesting case for future studies of the efficiency of grain growth process in these stages.
\item We also explore the effects of backwarming. The analysis shows that the envelope geometry highly affects the disk temperature. In the collapsing envelope model, the effect is weak, but if a power law envelope is used, the effect is more obvious.
\end{itemize}
Future high sensitivity data will be needed to allow us to conclusively prove whether there are spectral index variations between the disk and the envelope. Moreover, study of a larger sample of Class I sources in different star forming regions is important to understand how general this process is for grain growth.

\begin{acknowledgements}
      C.A.G. acknowledges support from CONICYT-Becas Chile (grant 72160297).
      P.C, J.P and L.S acknowledge the financial support of the European Research Council (ERC; project 320620). L.T. acknowledges the financial support of the Italian Ministero dell’Istruzione, Università e Ricerca through the grant Progetti Premiali 2012 -- iALMA (CUP C52I13000140001), and by the Deutsche Forschungs-gemeinschaft (DFG, German Research Foundation) -- Ref no. FOR 2634/1 TE 1024/1--1. MT has been supported by the DISCSIM project, grant agreement 341137 funded by the European Research Council under ERC-2013-ADG. A.M. acknowledges an ESO Fellowship.
\end{acknowledgements}

%
%






   
  



\bibliographystyle{aa}

\appendix
\begin{appendices}
\section{Error estimate of the spectral index of the observed flux densities}
The spectral index of the observed flux densities $\alpha_\mathrm{mm}$ can be approximated using the flux density at two wavelengths. In this appendix, we discuss the error propagation from the observational uncertainty to the deduced $\alpha_\mathrm{mm}$ value. Let $\mathrm{F_{1}}$ and $\mathrm{F}_{2}$ be the flux density at frequencies $\nu_{1}$ and $\nu_{2}$ , $\alpha_\mathrm{mm}$ can be expressed as in Equation (1):
	\begin{equation}
	\alpha_\mathrm{mm} = \frac{\ln \mathrm{F_{1}}-\ln \mathrm{F}_{2}}{\ln \mathrm{\nu}_{1}-\ln \mathrm{\nu}_{2}} .
	\end{equation}
We assume that the fluxes $\mathrm{F_{1}}$ and $\mathrm{F}_{2}$ are independent and that $\sigma_{F1}$ and $\sigma_{F2}$ are their standard deviations; using the error propagation we obtain
\begin{equation}
\sigma_\mathrm{\alpha}^{2} = \Bigg| \frac{\partial\alpha}{\partial\mathrm{F_{1}}} \Bigg|^{2} \sigma_{F1}^{2}  + \Bigg|\frac{\partial\alpha}{\partial\mathrm{F_{2}}}\Bigg|^{2}\sigma_{F2}^{2} \,
\end{equation}
Taking the partial derivative of Equation (A.1), we obtain
\begin{equation}
\frac{\partial\alpha}{\partial \mathrm{F_{1}}} = \frac{1}{(\mathrm{ln}\mathrm{\nu}_{1} - \mathrm{ln}\mathrm{\nu}_{2})\mathrm{F_{1}}}, \,
\end{equation}
\begin{equation}
\frac{\partial\alpha}{\partial\mathrm{F_{2}}} = \frac{1}{(\mathrm{ln}\mathrm{\nu}_{1} - \mathrm{ln}\mathrm{\nu}_{2}) \mathrm{F_{2}}}. \,
\end{equation}
Substituting equation A.3 and A.4 in equation A.2, the
uncertainty of the derived $\alpha_\mathrm{mm}$ is then:
\begin{equation}
\sigma_\mathrm{\alpha}^{2} = \Bigg ( \frac{1}{\mathrm{ln}\mathrm{\nu}_{1} - \mathrm{ln}\mathrm{\nu}_{2}} \Bigg)^{2}  \Bigg ( \frac{\sigma_{F1}^{2}}{\mathrm{F_{1}^{2}}}+ \frac{\sigma_{F2}^{2}}{\mathrm{F_{2}^{2}}}\Bigg). \,
\end{equation}

\section{$emcee$ implementation}
To compute the posterior distribution for all the free parameters, we use a variant of the Markov Chain Monte Carlo (MCMC) \citep{Mackay2003,Press2007} algorithm, which is widely known and efficient in finding a global maximum for a range of posteriors. We follow the affine-invariant ensemble sampler for MCMC by \citet{Goodman2010}, which basically transforms highly anisotropic and difficult-to-be-sampled multivariate posterior probability distribution function (PDFs) into isotropic Gaussians. The immediate advantage is that it is possible to simultaneously run many Markow chains ($walkers$) that will interact in order to converge to the maximum of the posterior.

This algorithm involves an ensemble \(S=\left\{X_k\right\}\) of simultaneously evolving \(K\) walkers, where the transition distribution for each walker is based on the current position of the other \(K-1\) walkers belonging to the complementary ensemble \(S_k=\left\{X_j,\ \forall j\ne k\right\}\). The position of a walker \(X_k(t)\) is updated as follows: 
\begin{equation}
X_k(t+1) = X_j + Z(X_k(t)-X_j),
\end{equation}
where \(X_j \in S_k\) and \(Z\) is a random variable drawn from a distribution that does not depend on the covariances between the parameters.

In this study we adopted an ensemble of 400 walkers, and let MCMC evolve for an initial burn--in phase. The burn--in phase is needed to allow MCMC to perform a consistent sampling of the space of parameters and to find the posterior maximum. To achieve the posterior maximum is needed to introduce the term: autocorrelation-time\footnote{Note: The longer the autocorrelation time, the larger the number of the samples we must generate to obtain the desired sampling of the posterior PDF.}, which is a direct measurement of the number of the posterior PDF evaluations needed to produce independent samples of the target density. 
For the analysis of Per-emb-50, 750 burn-in steps were performed to achieve convergence. Fig. B.1 presents a staircase plot, using the Python module $corner$ by \citet{corner}, showing the marginalized and bi-variate probability distributions resulting from the fit for Per-emb-50.

\begin{figure*}
\centering
\includegraphics[width=12cm]{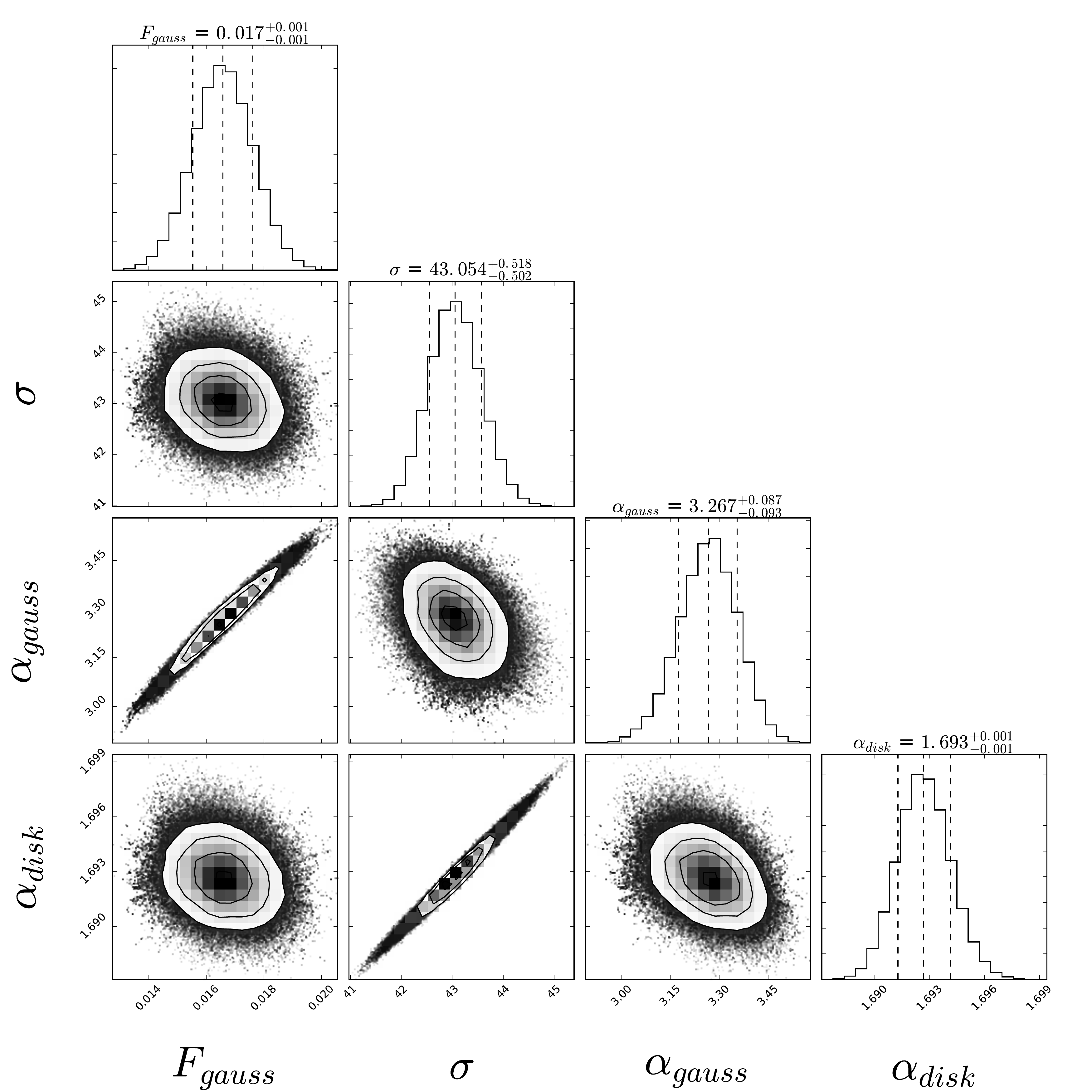}
\caption{Representation of the MCMC results for Per-emb-50. On the top diagonal, the 1D histograms are the marginalized distributions of the fitted parameters; the vertical dashed lines represent (from left to right) the 16th, the 50th, and the 84th percentiles. The 2D density plots represent the bi-variate distributions for each pair of parameters, with one dot representing one sample. The plot shows the posterior sampling provided by 1000 steps of the 400-walkers chain (750 burn-in steps were performed to achieve convergence).}
\label{appfig}
\end{figure*}
\section{Full Radiative transfer models}
The models presented in this appendix were created using a simple python module to set up RADMC-3D for disk plus envelope systems, 
{\ttfamily SimpleDiskEnv}\footnote{\url{https://gitlab.mpcdf.mpg.de/szucs/SimpleDiskEnv}}. Fig. C.1. show the best 36 models from a total of 288 for each maximum grain size (0.1,50,100,300,1000 $\mu$m).

\begin{figure*}
\centering
\includegraphics[width=6.55cm]{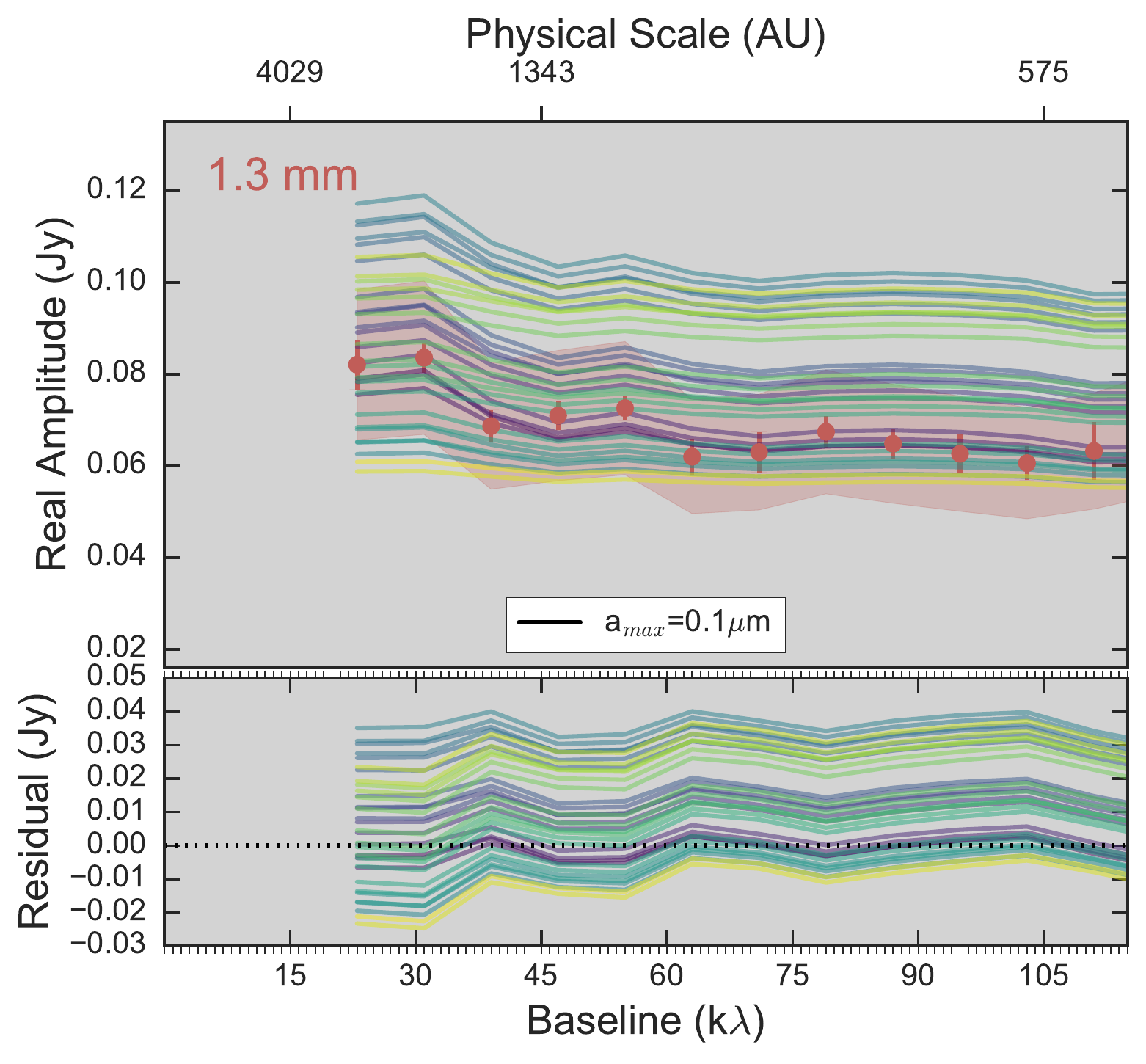}
\includegraphics[width=11.4cm]{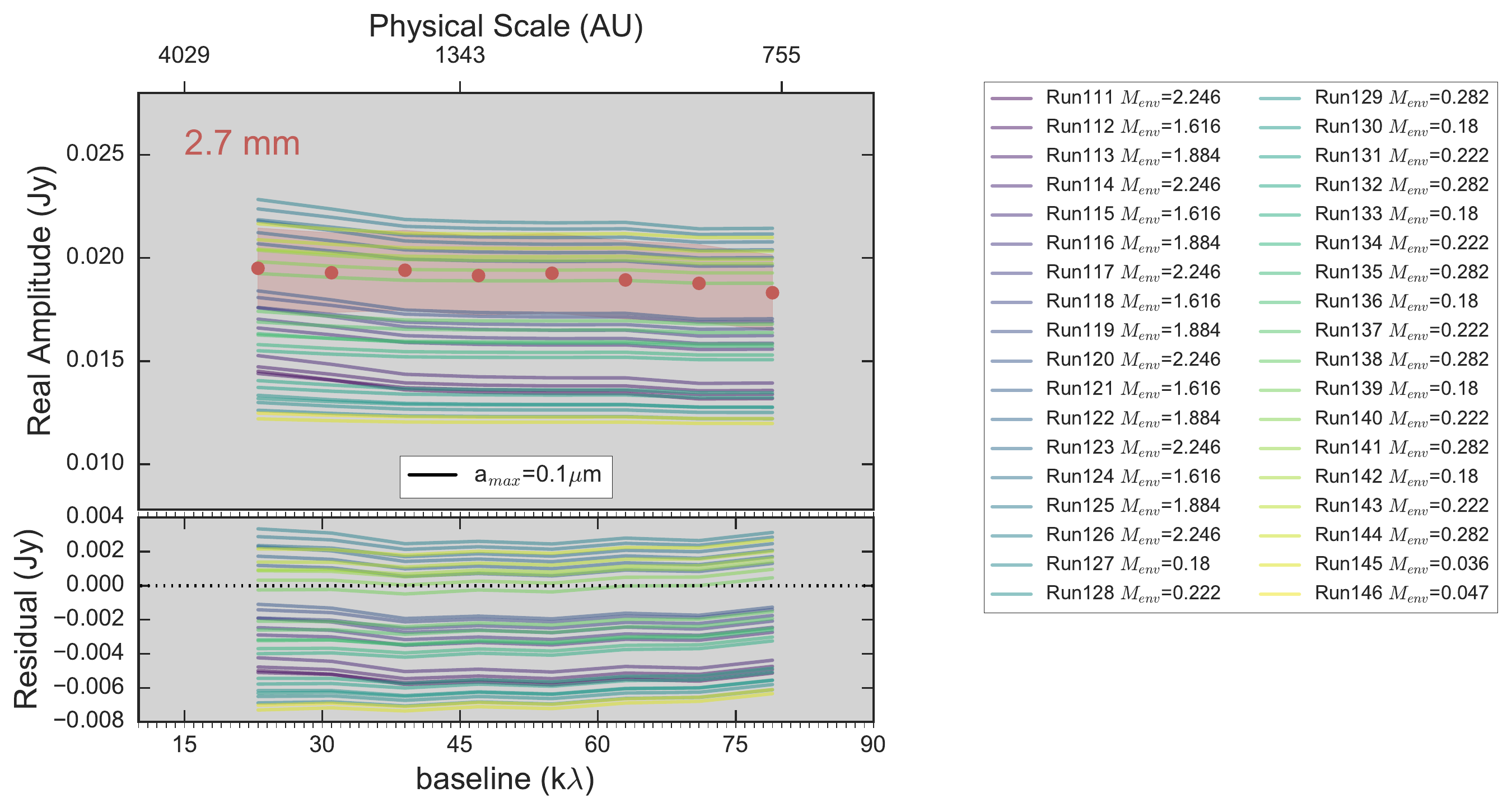}
\includegraphics[width=6.55cm]{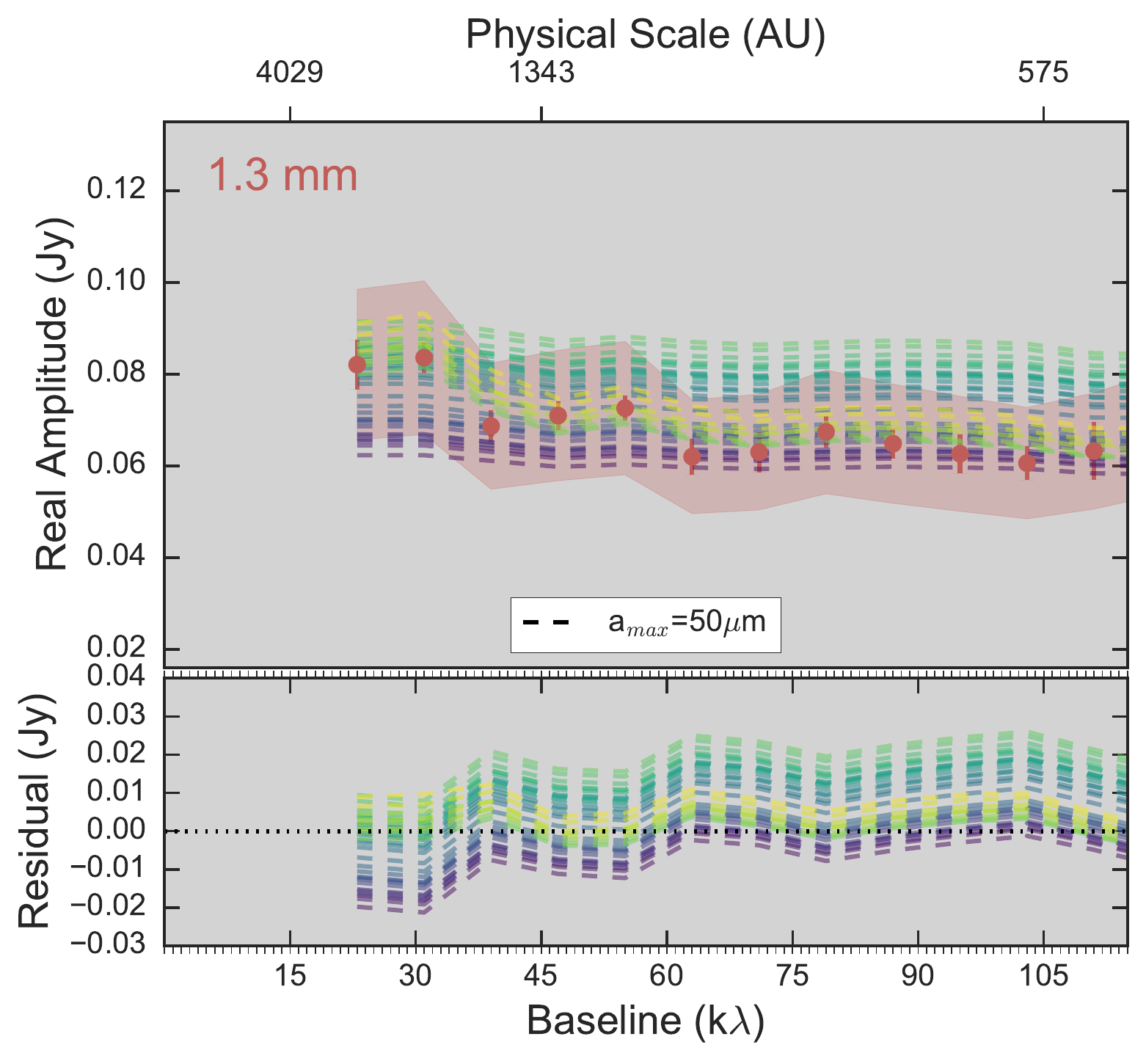}
\includegraphics[width=11.4cm]{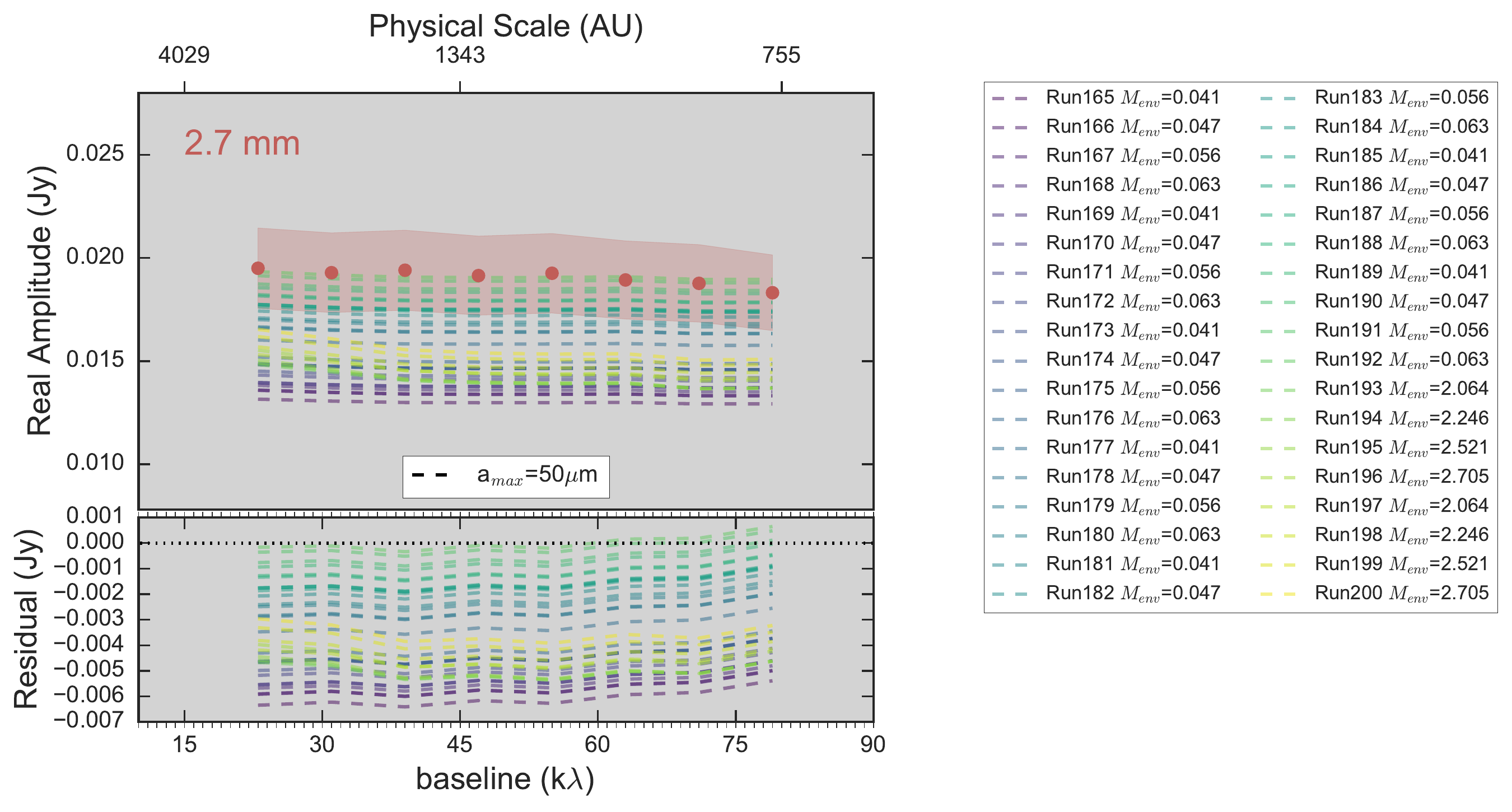}
\includegraphics[width=6.55cm]{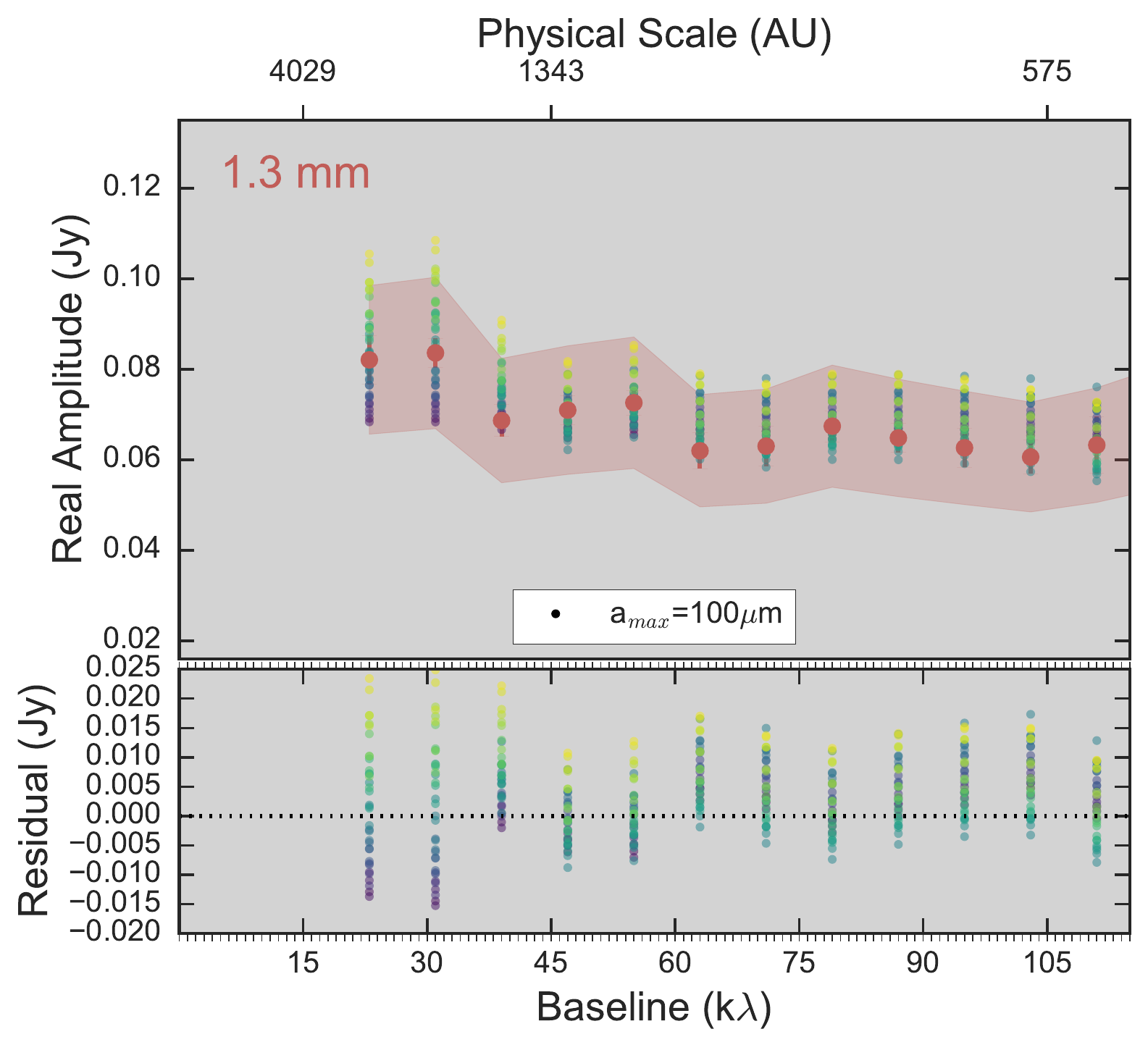}
\includegraphics[width=11.4cm]{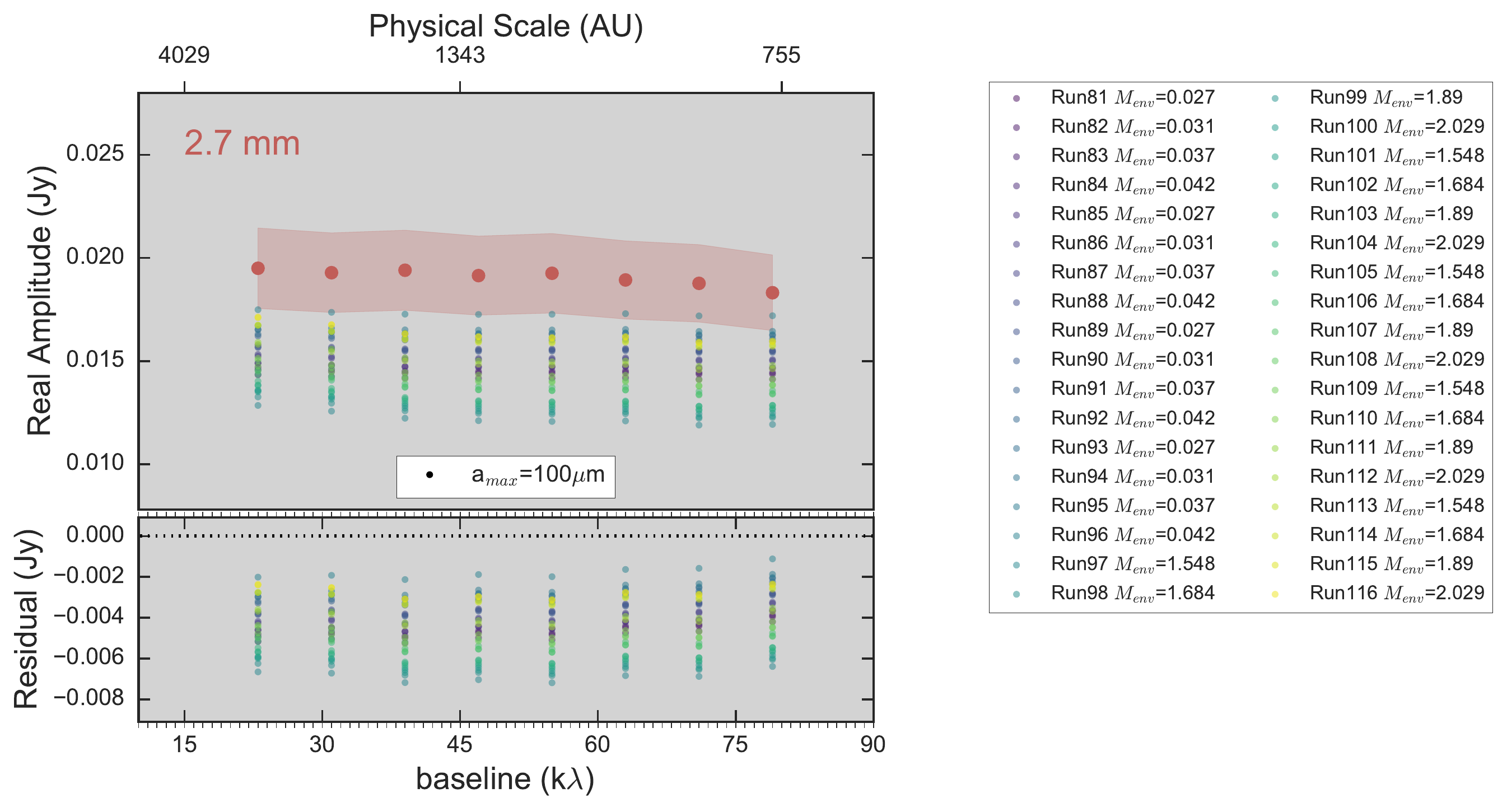}
\includegraphics[width=6.55cm]{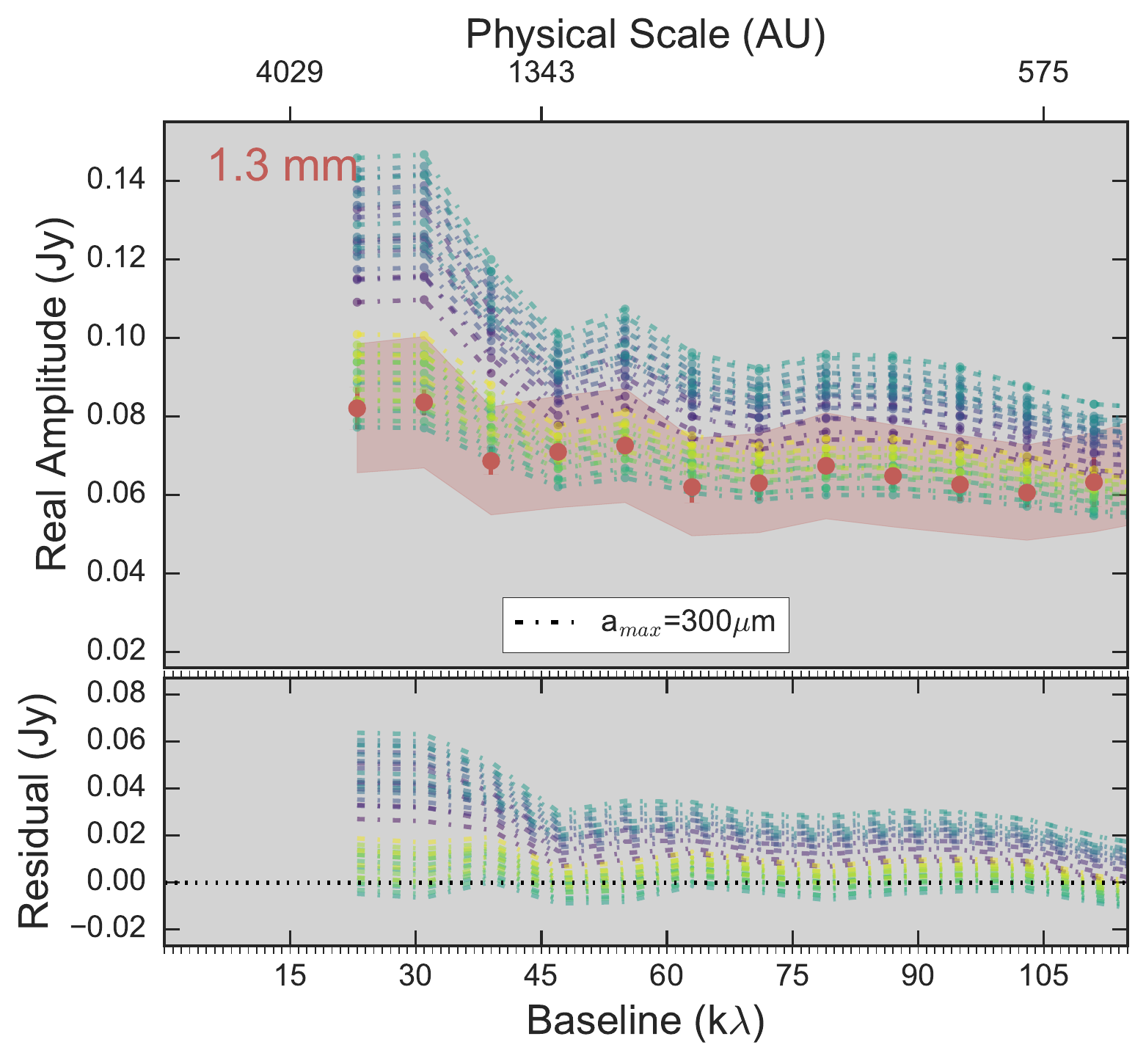}
\includegraphics[width=11.4cm]{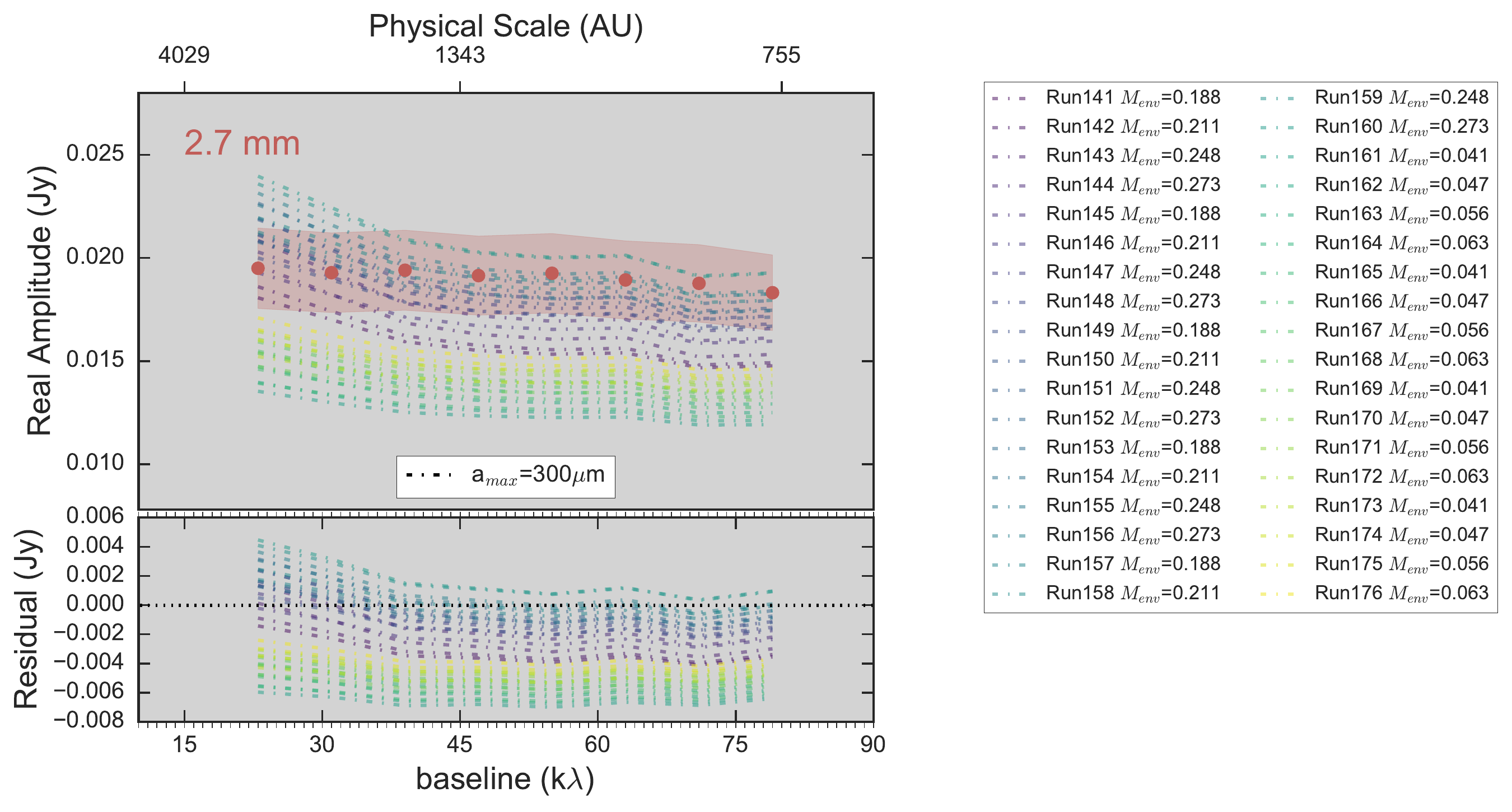}
\label{144models}
\end{figure*}
\begin{figure*}
\centering
\includegraphics[width=6.55cm]{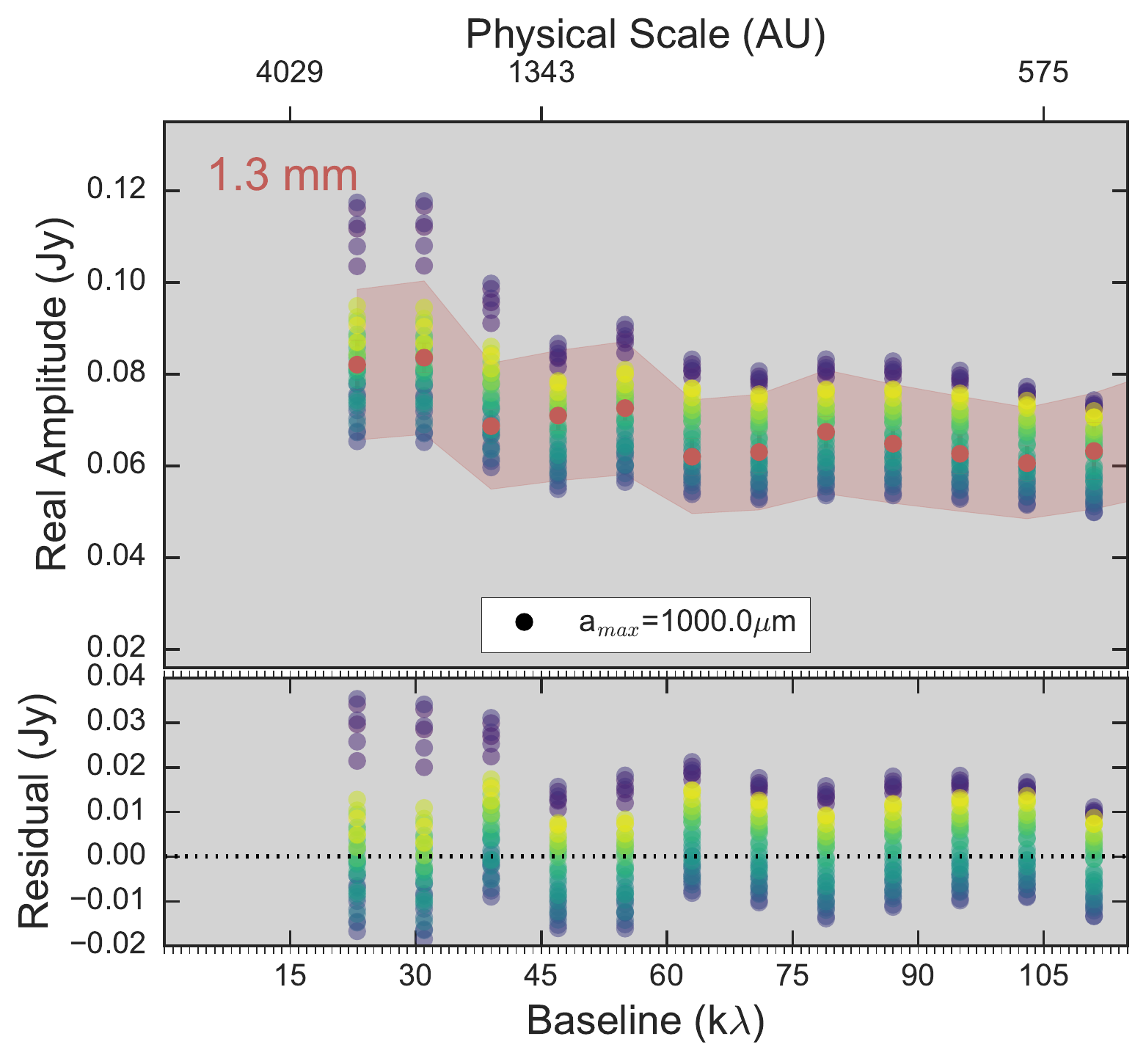}
\includegraphics[width=11.4cm]{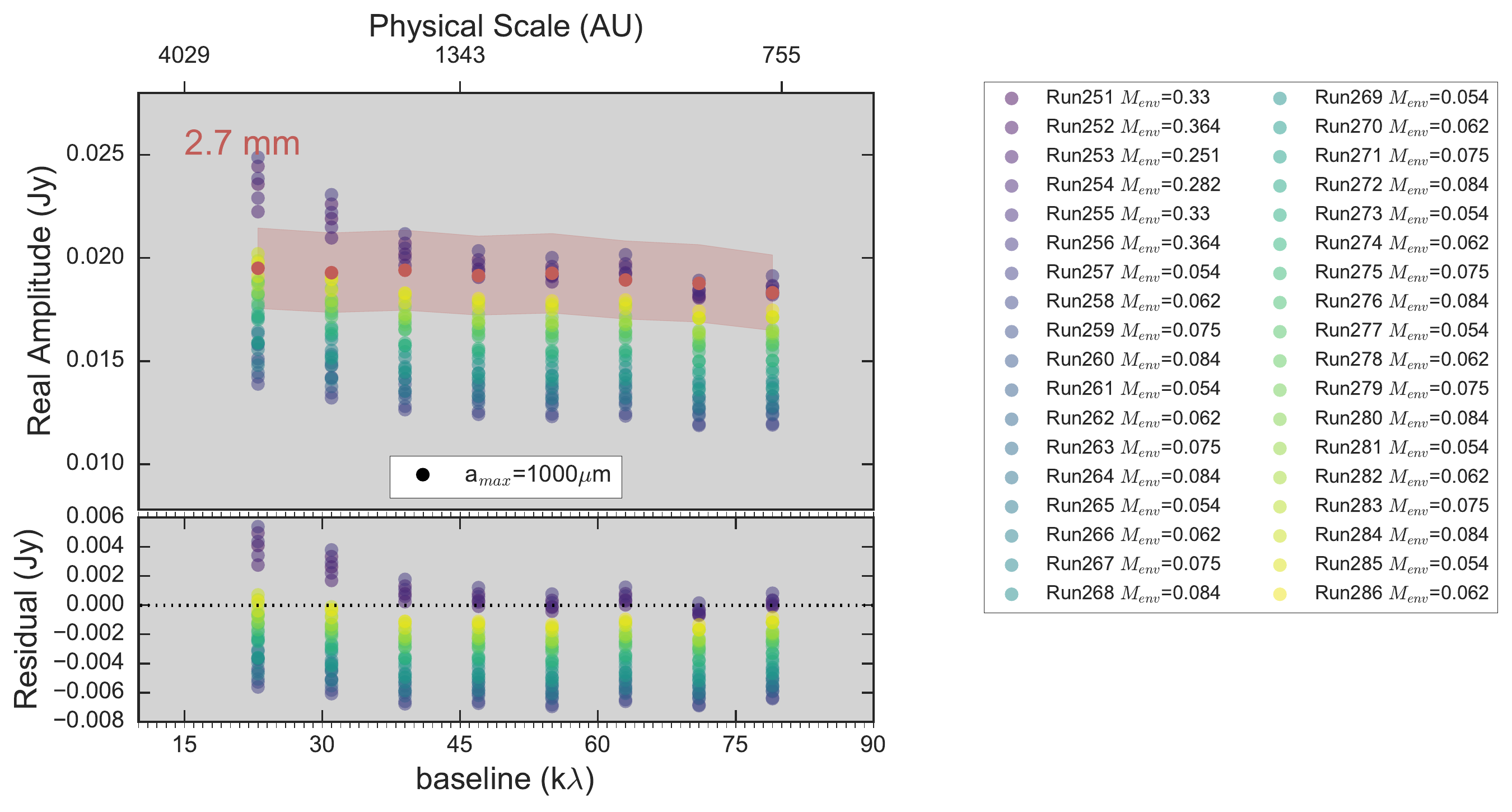}

\caption{Full radiative transfer models for different $a_{max}$. The name of the model and derived envelope mass are in the right panel. The color gradient represent the $\chi^{2}$ from lower (blue) to high values (yellow), that were used only as reference. After visual inspection, we choose the best models from the green area.}
\label{144models}
\end{figure*}
\section{Backwarming effect}
We studied the net effect of the envelope on the disk temperature using a RADMC-3D toy model of a Class I protostar. 
As mentioned in \citet{Butner1994}, the envelope can have an important backwarming effect on the disk, affecting the outer edges of the disk with a flat temperature distribution. \\
To probe this effect, we first modeled a disk of 25~AU without an envelope and with a distribution of dust grains in the disk with a maximum size $a_{max}^{disk}$=1 cm. Then we add a 1.3 M$_{\odot}$ envelope, with a \citet{Tafalla2002} density profile and grain sizes with $a_{max}^{env}$=100 $\mu$m. The inner edge of the envelope and the outer radius of the disk are the same. To compare and quantify the effect, we model a disk with the same characteristics but with a density profile of a collapsing envelope defined by \citet{Ulrich1976}. 
Fig D.1. shows the temperature structure (in cylindrical coordinates) for both of these cases. In the left panels (disk only) we can see that the outer regions of the disk are around 20--30\,K. In the right upper panel (disk+envelope) using the \citeauthor{Ulrich1976} envelope structure, the temperature increases to 40--60\,K. In the case of the model with a \citeauthor{Tafalla2002} envelope structure the effect is quite strong, reaching disk outer temperatures of 120--140\,K.

\begin{figure*}[h!]
\centering
\includegraphics[width=9cm]{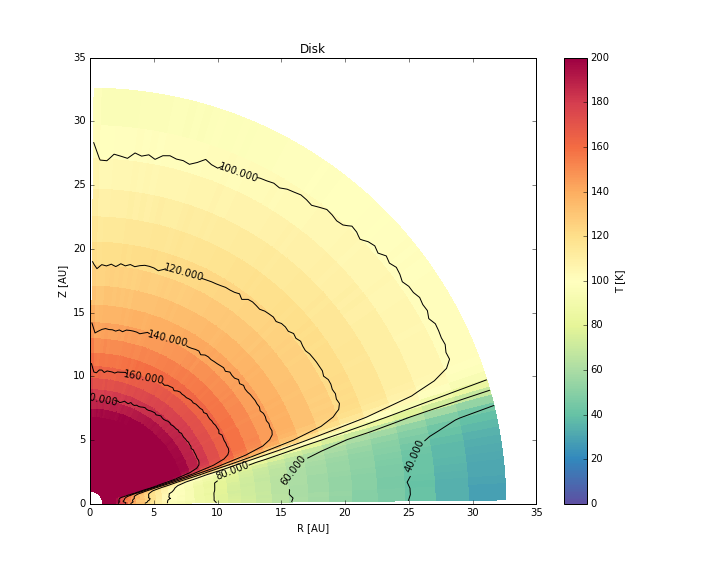}
\includegraphics[width=9cm]{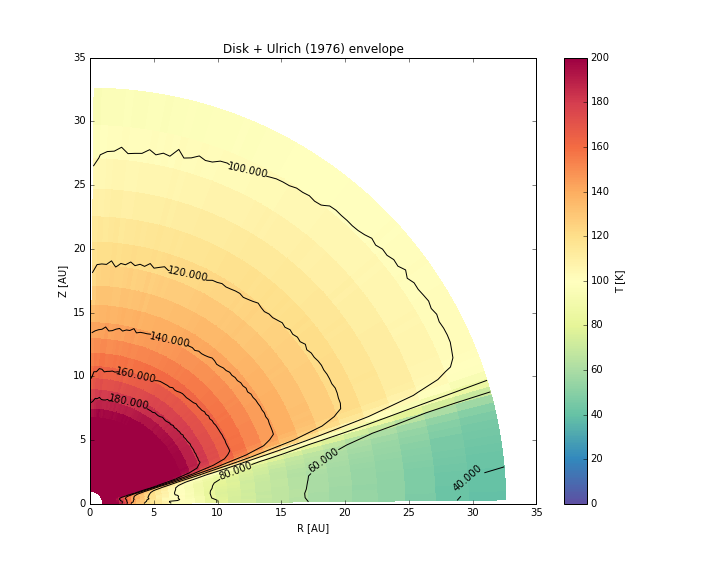}
\includegraphics[width=9cm]{disk}
\includegraphics[width=9cm]{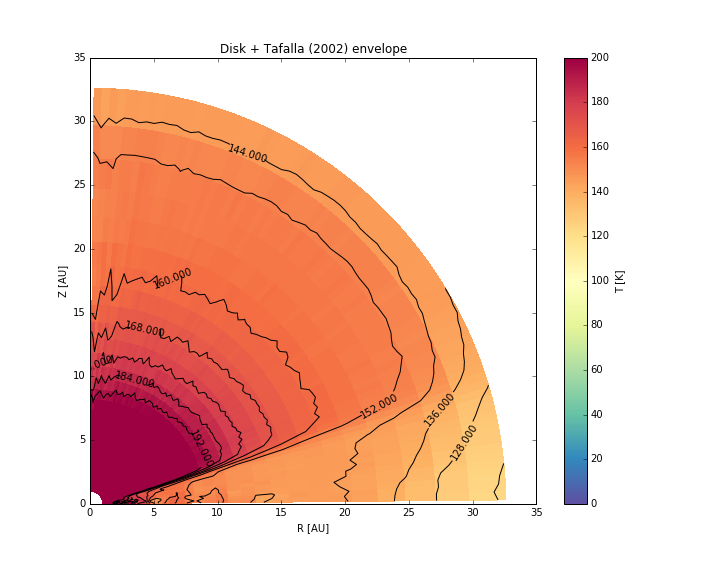}
\caption{Temperature structure in cylindrical coordinates of two cases: (top left panel) 25~AU disk with $a_{max}^{disk}$=1 cm and (top right panel) a 25~AU disk with a 1.3 M$_{\odot}$ \citeauthor{Ulrich1976} envelope structure and grain sizes with $a_{max}^{env}$=100 $\mu$m. (Bottom right panel) is the case of a 1.3 M$_{\odot}$ \citeauthor{Tafalla2002} envelope profile with $a_{max}^{env}$=100 $\mu$m heating the 25~AU disk. 2D temperature contours are presented in black lines.}
\label{appfig}
\end{figure*}

\end{appendices}

\end{document}